\documentclass[9pt, oneside]{article}   	
\usepackage{geometry}                		
\usepackage{graphicx}				
\usepackage{amsmath}
\usepackage{amssymb}
\usepackage{dsfont}
\usepackage{algorithm}
\usepackage[noend]{algpseudocode}
\usepackage{hyperref}
\usepackage{doi}
\usepackage{pf2}
\usepackage{todonotes}
\usepackage[most]{tcolorbox}

\tcbset{
    frame code={}
    center title,
    left=0pt,
    right=0pt,
    top=0pt,
    bottom=0pt,
    colback=gray!45,
    colframe=black,
    width=\dimexpr\textwidth\relax,
    enlarge left by=0mm,
    boxsep=5pt,
    arc=5pt,outer arc=5pt,
    }

\newcommand{\defeq}{\overset{\mathrm{def}}{=}}

\newcommand{\anyleadsto}[1]{\overset{#1}\leadsto}
\newcommand{\logleadsto}{\overset{\textit{log}}\leadsto}

\title{2P-BFT-Log: 2-Phase Single-Author Append-Only Log for Adversarial Environments}
\author{Erick Lavoie \\
University of Basel \\
\href{mailto:erick.lavoie@unibas.ch}{erick.lavoie@unibas.ch}}
\date{July 28, 2023}							

\begin{document}
\maketitle

\begin{abstract}
Replicated append-only logs sequentially order messages from the same author such that their ordering can be eventually recovered even with out-of-order and unreliable dissemination of individual messages. They are widely used for implementing replicated services in both clouds and peer-to-peer environments because they provide simple and efficient incremental reconciliation. However, existing designs of replicated append-only logs assume replicas faithfully maintain the sequential properties of logs and do not provide eventual consistency when malicious participants fork their logs by disseminating different messages to different replicas for the same index, which may result in partitioning of replicas according to which branch was first replicated.
 
In this paper, we present \textit{2P-BFT-Log}, a two-phase replicated append-only log that provides eventual consistency in the presence of forks from malicious participants such that all correct replicas will eventually agree either on the most recent message of a valid log (first phase) or on the earliest point at which a fork occurred as well as on an irrefutable proof that it happened (second phase). We provide definitions, algorithms, and proofs of the key properties of the design, and explain one way to implement the design onto Git, an eventually consistent replicated database originally designed for distributed version control.

Our design enables correct replicas to faithfully implement the happens-before relationship first introduced by Lamport that underpins most existing distributed algorithms, with eventual detection of forks from malicious participants to exclude the latter from further progress. This opens the door to adaptations of existing distributed algorithms to a cheaper \textit{detect} and \textit{repair} paradigm, rather than the more common and expensive systematic \textit{prevention} of incorrect behaviour.
\end{abstract}

\section{Introduction}
\label{sec:introduction}

An append-only log, as used in eventually consistent replicated databases in clouds~\cite{kappa-architecture,decandia2007dynamo} or peer-to-peer systems~\cite{kermarrec2020gossiping,hypercore-website}, is a totally ordered replicated list of messages such that each index has at most one unique message associated to it and indexes are associated with new messages from lowest to greatest. In trusted environments, such as clouds, messages are associated to globally assigned process identifiers, while in peer-to-peer environments, messages are usually associated to authors as represented by the public key of a public-private key pair. In this paper we focus on append-only logs using messages associated to authors.

In the simplest implementation of an append-only log, each new message from the same author is stored at the end of an ever growing list. When two replicas reconcile the state of the log associated to a given author, they compare the index of the latest messages they have. When one replica's index is larger, the replica with the most recent messages sends the missing messages to the other~\cite{vanrenesse2008scuttlebutt,kermarrec2020gossiping}. 

To prevent tampering and ensure authenticity, some peer-to-peer systems~\cite{kermarrec2020gossiping,hypercore-website} use cryptographic hashes to link a log's new message to its immediate predecessor. However, this strategy does not prevent a malicious author from creating two \textit{concurrent} messages for the same index~\cite{lavoie2023gocledger,hypercore-handling-conflicts,hypercore-split-resolution-dep} thereby \textit{forking} their log and turning it into a tree. Cryptographic hashes are therefore not sufficient to sequentially order the messages of a log.

Also, append-only logs, as currently implemented in some peer-to-peer systems~\cite{kermarrec2020gossiping,lavoie2023gocledger,hypercore-handling-conflicts,hypercore-split-resolution-dep},  cannot tolerate malicious participants~\cite{Kleppmann2022byzantine,jacob2022bft-crdt} because once a replica has replicated one of the concurrent branches of a forked log, the replica will consider any other branch updates as invalid. This results in a partitioning of correct replicas, preventing convergence.

In this paper, we present the design of a two-phase Byzantine Fault-Tolerant append-only log, \textit{2P-BFT-Log}, which provides eventual consistency in the presence of malicious authors that may intentionally fork their log. The key idea and novel contribution in our design is to introduce a second \textit{shrinking} phase after a fork has been discovered, such that all correct replicas agree on the earliest fork that has been observed so far. An irrefutable proof, in the form of at least two signed messages from the malicious author that have the same predecessor message, is replicated between correct replicas to establish the earliest fork point. In the context of decentralized accounting~\cite{lavoie2023gocledger}, for example, our design provides two key properties for both correct and malicious authors: for correct authors, the total ordering of messages guarantees valid updates always maintain non-negative balances for accounts, while for malicious authors, it provides eventual detection of potential double-spending because double-spending requires concurrent messages.

Moreover, the switch of forked logs to an explicit and separate shrinking phase eventually prevents malicious authors from being able to extend their log with correct replicas: as soon a correct replica finds a fork proof, malicious authors have a limited window of opportunity to propagate new forks to an ever shrinking set of correct replicas that have not yet received the fork proof. After that opportunity is over, the log of a malicious author is \textit{dead} as it cannot be extended anymore.

In the rest of this paper, we first introduce relevant background (Section~\ref{sec:background}), then the design of \textit{2P-BFT-Log} (Section~\ref{sec:design}), an implementation based on Git (Section~\ref{sec:git-implementation}),  proofs for convergence and key properties (Section~\ref{sec:proofs}), a review of related work (Section~\ref{sec:rel-work}), and we finally conclude with a brief recap and directions for future work (Section~\ref{sec:conclusion}).

\section{Background}
\label{sec:background}

In this section, we present relevant background, including some of its limitations, that motivates our design.

\subsection{Byzantine Processes}

A Byzantine process~\cite{lamport2019byzantinegenerals} is a process that may deviate arbitrarily from the behaviour otherwise expected of correct processes, i.e. it does not execute the algorithms faithfully. For example, it may omit to send some messages, send invalid messages, try impersonating other processes, broadcast inconsistent messages to different processes, etc. In general, it is not possible to detect Byzantine processes until they misbehave, and not all misbehaviour can be reliably detected. However, using cryptographic signatures it is possible to remove the ability of Byzantine processes to impersonate or tamper with messages they relay from other correct processes~\cite{lamport2019byzantinegenerals}, and detect inconsistent messages broadcast to different correct processes by comparing the messages received by correct processes.

\subsection{Happens-Before Relationship}
\label{sec:background:happens-before}

The \textit{happens-before} relationship~\cite{lamport2019time} is a causal ordering of events happening on concurrent processes. We say that an event $e$ happens before $e'$, written $e \leadsto e'$, if there is a sequence of causes and effects that connects $e$ to $e'$. If no sequence of causes and effects connect $e$ and $e'$, i.e. $e \not\leadsto e'$ and $e' \not\leadsto e$, then we say that $e$ and $e'$ are concurrent, written $e \parallel e'$.

More formally, we assume that all processes locally execute sequentially so that there exists a total order $<_e$ on all events happening within that process. Then, for two events $e$ and $e'$ happening on processes $p$ and $p'$, $e \leadsto e'$ if one of the following conditions is true, otherwise $e \not\leadsto e'$:
\begin{itemize}
	\item \textbf{local total order}: $p=p'$ and $e <_e e'$, i.e. $e$ happened before $e'$ within the same process;
	\item \textbf{interprocess communication}: $e$ happened before communication between $p$ and $p'$, for example message $m$ was sent from $p$ to $p'$, and $e'$ happened after communication between $p$ and $p'$, for example message $m$ was received on $p'$\footnote{$p$ may send a message to itself, therefore $p$ may be the same as $p'$.};
	\item \textbf{transitivity}: there exists $e''$ such that $e \leadsto e''$ and $e'' \leadsto e'$.
\end{itemize}

Note that the assumption of a local total order is easy to satisfy on correct processes but impossible to enforce on Byzantine processes. Also, the communication between processes could actually be a full reconciliation protocol (ex: \cite{kleppmann2020bec}) rather than a single message sent. 

\subsection{Hash Graphs}
\label{sec:background:hash-graph}

A hash graph~\cite{jacob2022bft-crdt,Kleppmann2022byzantine} is a widely used implementation technique to encode the causal relationships of messages using cryptographic hashes. It is used, for example, to order Git commits~\cite{git-commit}, authenticate file updates~\cite{mahajan2011depot}, and design Byzantine fault-tolerant algorithms~\cite{maniatis2002secure-timeline-entanglement, cachin2011introduction,kleppmann2020bec, Kleppmann2022byzantine}. It is standard to assume that the cryptographic hashes of messages are unique,\footnote{More precisely, there is a negligible probability for such a collision.} i.e. there does not exist messages $M, M'$ such that $M \neq M'$ and $H(M) = H(M')$. By including the cryptographic hashes of all messages $M'$ on which $M$ depends within $M$, we can then encode the fact that $M$ happened before and may have caused $M'$. 

Assume that any message $M$ stores the set of messages it depends on in $M_\textit{deps}$. We say that given two valid messages $M$ and $M'$, $M$ is a \textit{predecessor} of $M'$ if and only if one of the following conditions is true:
\begin{itemize}
	\item \textbf{direct dependency}: $H(M) \in M'_\textit{deps}$;
	\item \textbf{transitivity}: there exists $M''$ such that $M$ is a predecessor of $M''$ and $M''$ is a predecessor of $M'$.
\end{itemize}

Similarly, a successor of $M$ is a message $M'$ such that $M$ is a predecessor of $M'$. By construction, a hash graph is a directed acyclic graph, i.e. a successor $M'$ of $M$ cannot be a dependency of $M$ because the successors between $M'$ and $M$ would have to be created before $H(M')$ is computed which would require $H(M)$ to be computed first, which in turn requires knowing $H(M')$, introducing a circular dependency.\footnote{In practice, one could try a large number of possibility until a consistent cycle was found but the probability of finding one is negligible.}

It is customary to assign an explicit author to messages of hash graphs. In general, the authors of hash graph messages are independent of the underlying processes that generate messages: a process may sequentially generate messages from a single author that appear concurrent in the hash graph, i.e. two messages with the same predecessor in their dependencies. Therefore, although the cryptographic hashes encode causal relationships and authors may be seen as a reification of process identifiers of the happens-before relationship, the analogy is incomplete because a hash graph does not enforce a total ordering of messages from the same author.

\subsection{Conflict-Free Replicated Data Types}
\label{sec:background:crdt}

Conflict-Free Replicated Data Types (CRDTs)~\cite{Shapiro2011CRDTs}  are mutable replicated objects designed with constraints on concurrent modifications such that all replicas\footnote{When discussing CRDTs, we use \textit{replica} to designate the process that executes the algorithm, to emphasize the idea that the purpose of the process is to replicate the object.} are always guaranteed to converge to the same state \textit{eventually}. More precisely, they provide \textit{strong eventual consistency}, defined as:
\begin{itemize}
	\item \textbf{eventual update}: If an update is applied by a correct replica, then all correct replicas will eventually apply that update;
	\item \textbf{convergence}: Any two correct replicas that have applied the same set of updates are in the same state (even if updates were applied in different orders).
\end{itemize} 

State-based CRDTs immediately update their local state and later send their new state to other replicas. The eventual update property is obtained by assuming correct replicas are transitively connected and new states are eventually received and merged by neighbours. Convergence is obtained by organizing all possible states of objects into a partially ordered set, defining a deterministic merge function for any two possible states that computes the least upper bound within that partial order, and constraining update operations such that they only produce a new state that is equal or larger within the partial order. Therefore, the convergence proofs for state-based CRDTs are relatively easy and do not require reasoning about possible orderings of updates.

Operation-based CRDTs define conflict resolution rules for all possible concurrent updates but require reliable delivery of all updates to all replicas. The convergence proofs for operation-based CRDTs are somewhat more involved and require reasoning about causal ordering of updates. In the rest of this paper, we only concern ourselves with state-based CRDTs.

\subsection{Byzantine State-Based CRDTs}
\label{sec:background:byzantine-state-crdt}

Most of the literature on CRDTs~\cite{crdt-website} assumes that processes are non-Byzantine and many CRDT designs break down in the presence of Byzantine replicas~\cite{Kleppmann2022byzantine}. State-based CRDTs however do not require adjustments~\cite{jacob2022bft-crdt}. Beside producing invalid states that can be easily rejected by correct replicas, Byzantine processes are left with two possibilities:
\begin{itemize}
	\item \textit{omission} of sending some state updates to certain replicas;
	\item \textit{equivocation}, i.e. sending different state updates to different replicas.
\end{itemize}

The second case is actually a variation of the first one, in that two different state updates are sent to different strict subsets of correct replicas. As long as at least one correct replica receives each state update, the \textit{eventual update} property (Section~\ref{sec:background:crdt}) guarantees all correct replicas will apply it and the \textit{convergence} property guarantees that all correct replicas converge to the same state. The only issue remaining is to define what are valid states and only merge those, which is not significantly more involved than designing CRDTs in a non-Byzantine context, making their description and proofs quite accessible.

\subsection{Self-Certification}
\label{sec:background:self-certification}

Self-certification~\cite{mazieres2000selfcertifyingfs} is a property of a combination of reference and the referred data such that they can be compared to ensure authenticity. We use two variations  in this paper:
\begin{itemize}
	\item a message $m$ retrieved using its hash $H(m)$ is self-certifying because the retriever can compare the hash of the data obtained to ensure it is equal to the hash they requested;
	\item the messages of a cryptographically signed append-only log retrieved using the public key of the author of messages can be checked against the public key. 
\end{itemize}

When used with public-private keys, self-certification does not however by itself guarantees that a public key actually belongs to the author it claims to represent. It does however, removes the ability from Byzantine authors to impersonate another author without the need for a third-party to manage keys.

We now combine these ideas into a new append-only log design.

\section{2P-BFT-Logs Design}
\label{sec:design}

A \textit{2P-BFT-Log} solves the problem of  totally ordering messages from a given author, similar to an append-only log, while providing eventual consistency and fault-detection in the presence of Byzantine authors that create concurrent messages.
Our key and novel contribution is to correctly handle the possibility of \textit{multiple forks} on the same log by implementing a distinct \textit{shrinking phase}. We present our design progressively in three steps: first we discuss the underlying grow-only set of immutable messages (Section~\ref{sec:design:message}), then a single mutable log made of messages (Section~\ref{sec:design:log}), and finally sets of logs (Section~\ref{sec:design:frontier}).

\subsection{System Model}

We assume there is an arbitrary large set of replicas, among which there is subset of at least two that are correct, and the rest is Byzantine, i.e. can deviate arbitrarily from our algorithms. Every correct replica is connected to at least one other correct replica such that, through that replica, they receive updates from all other correct replica, i.e. correct replicas from a connected component. We further assume that every state update of a correct replica is eventually received by all other replicas: this is a weaker assumption than reliable broadcast because the update may be received indirectly through another correct replica and might have been merged with another state update before being received. This last assumption can be captured theoretically by assuming every replica send their latest state infinitely often to their immediate neighbour in the connection graph so that at least one transmission eventually succeed.

In addition, we have a set of authors that use the replicas to propagate their messages. Authors may be correct, in which case they only create valid messages that follow a sequential ordering according to the hash graph we will explain shortly, or they may be Byzantine in which case they may create both invalid messages as well as valid messages that do not follow a sequential ordering, i.e. multiple concurrent messages with the same predecessor in the hash graph. Correct replicas will always reject invalid messages but Byzantine replicas may do anything.

We assume both Byzantine authors and replicas do not have access to the private keys of correct authors, therefore they cannot forge messages in their name. We make no assumption on whether Byzantine authors or replicas share private keys.

\subsection{Message Graph}
\label{sec:design:message}

The message graph is the underlying backbone of logs. It explicitly encodes the causal dependencies between messages, similar to the commit graph in Git~\cite{git-commit}.  We do not discuss how messages are replicated: there already exists reconciliation protocols that would be sufficient for the task (ex: \cite{kleppmann2020bec}), our only requirement is that all dependencies of a message should also be replicated to enable validation.

Our message graph is similar to a hash graph that includes cryptographic signatures from authors (Section~\ref{sec:background:hash-graph}). The signatures prevent Byzantine authors from impersonating other authors or tampering with messages during replication. In contrast to some hash graph designs~\cite{kleppmann2020bec, Kleppmann2022byzantine, jacob2022bft-crdt}, but similar to entangled timelines~\cite{maniatis2002secure-timeline-entanglement}, we restrict valid messages to at most one message dependency from the same author: this constrains the message graph of any given author to a sequence, if correct, or a tree, if Byzantine. Messages follow the schema of Table~\ref{tb:message-schema}. In contrast to the usual definition for append-only logs~\cite{kermarrec2020gossiping}, we also encode dependencies to messages from other authors so that the full causal history of a message can easily be recovered. To ease the presentation of later relations and algorithms we use a separate field $M_\textit{prev}$ for the dependency from the same author, while all other dependencies are stored in $M_\textit{deps}$. 

\begin{table}[t]
\caption{Message schema.}
\label{tb:message-schema}
\begin{centering}
\begin{tabular}{lll}
\textbf{Field}         & \textbf{Type}     & \textbf{Purpose} \\ \hline
$M_\textit{author}$  & String (Public Key)           &  Author of the message.                \\
$M_\textit{prev}$      & MsgId or $\bot$    & Previous message from the same author, $\bot$ (null) if none.                 \\
$M_\textit{deps}$      & Set of MsgIds      & State of logs from other authors  on which it depends.            \\
$M_\textit{payload}$      & String &  Content of message, which typically represents an update.                \\
$M_\textit{signature}$ & String            & \begin{tabular}{@{}l@{}}
							      Signature of the concatenation of the previous fields using \\
							       \textit{author}'s private key.    
							       \end{tabular}            
\end{tabular}
\newline
\end{centering}
\end{table}

A message identifier for message $M$, abbreviated $\texttt{MsgId}(M)$, is computed by hashing the concatenation of all message fields, making the identifer and message pair self-certifying (Section~\ref{sec:background:self-certification}):

\begin{equation*}
\texttt{MsgId}(M) \defeq \texttt{hash}(M_{\textit{author}} \oplus M_{\textit{prev}} \oplus M_{\textit{idx}} \oplus M_{\textit{payload}} \oplus M_{\textit{deps}} \oplus M_{\textit{signature}})
\end{equation*}

\subsubsection{Validity Properties}
\label{sec:design:message:valid}

We validate messages as follows. Either there is no previous message from the same author, or there is at most one and it is valid: 

\begin{tcolorbox}
\begin{enumerate}
	\item[\textbf{M1}] \textit{(single previous dependency)}: either $\begin{cases}
		M_\textit{prev} = \bot; \\
		M_\textit{prev} = \texttt{MsgId}(M') : M' ~\text{exists and is valid}.
		\end{cases}$ \\
\end{enumerate}
\end{tcolorbox}

Previous messages are from the same author:
\begin{tcolorbox}
\begin{enumerate}
	\item[\textbf{M2}] \textit{(single author)}: if $M_\textit{prev} = \texttt{MsgId}(M')$ then $M'_\textit{author} = M_\textit{author}$.
\end{enumerate}
\end{tcolorbox}

All dependencies are from different authors and valid:
\begin{tcolorbox}
\begin{enumerate}
	\item[\textbf{M3}] \textit{(valid external dependencies)}: for all $\textit{msgId} \in M_\textit{deps}$, there exists a valid $M'$ such that $\texttt{MsgId}(M') = \textit{msgId}$ and $M'_\textit{author} \neq M_\textit{author}$.
\end{enumerate}
\end{tcolorbox}

There is at most one external dependency per author:
\begin{tcolorbox}
\begin{enumerate}
	\item[\textbf{M4}] \textit{(single author dependencies)}: for any possible \textit{author} not equal to $M_\textit{author}$, there exists at most one $\textit{msgId} \in M_\textit{deps}$, such that $\texttt{MsgId}(M') = \textit{msgId}$ and $M'_\textit{author} = \textit{author}$.
\end{enumerate}
\end{tcolorbox}

The signature is valid:
\begin{tcolorbox}
\begin{enumerate}
	\item[\textbf{M5}] \textit{(self-certifying)}: $M_\textit{signature}$ is consistent with $M_\textit{author}$.
\end{enumerate}
\end{tcolorbox}

There are no dependency cycle (the definition for successor is given later in Section~\ref{sec:design:writer-rel-queries}):
\begin{tcolorbox}
\begin{enumerate}
	\item[\textbf{M6}] \textit{(acyclic)}: $M_\textit{prev}$ does not reference a successor of $M$, i.e. ($M_\textit{prev} \neq M' : M \logleadsto M'$).
\end{enumerate}
\end{tcolorbox}

A standard cryptographic assumption is to assume the hash function used for $\texttt{MsgId}(M)$ does not have collisions, which prevents cycles. In practice, even if that possibility is non-null it suffices to only validate a message if all its predecessors and dependencies have been previously successfully validated. This way even if a successful hash cycle had been found by a Byzantine author, no correct replica will accept it.

\subsubsection{Relations and Queries on Messages}
\label{sec:design:queries}

We now present useful relationships and queries on messages in a message graph. These help make the presentation of algorithms  precise and concise, and they also provide formal semantics for Git+Bash commands used in our implementation (Section~\ref{sec:git-implementation:rel-queries}). All the following relations and queries are only defined for valid messages, i.e. messages that meet the properties of Section~\ref{sec:design:message:valid}.

The \textit{happens-before relationship} follows easily from its usual definition (Section~\ref{sec:background:happens-before}):

\begin{equation}
	M \leadsto M' \defeq \begin{cases}
		M = M'_\textit{prev} = \bot \\
		\texttt{MsgId}(M) = M'_\textit{prev} \\
		\texttt{MsgId}(M) \in  M'_\textit{deps} \\
		\exists M'' : M \leadsto M'' \leadsto M'
	\end{cases}
\end{equation}

From the happens-before relationship, we define a partial order on messages ($\leq_\mathcal{M}$) in which $M$ is smaller or equal to $M'$ if it is either equal or happens before $M'$:
\begin{equation}
	M \leq_\mathcal{M} M' \defeq M = M' \vee M \leadsto M' 
\end{equation}

It is easy to show that $\leq_\mathcal{M}$ is \textit{reflexive}, \textit{transitive}, and \textit{antisymmetric} and indeed forms a partial order, so we omit the proof for brevity. A partial ordering is useful to define CRDTs as we do later.

If neither $M$ or $M'$ is smaller or equal to the other, then they are concurrent ($\parallel_\mathcal{M}$):
\begin{equation}
	M \parallel_\mathcal{M} M' \defeq M \not\leq_\mathcal{M} M' \wedge M' ~\not\leq_\mathcal{M}~ M
\end{equation}

~\newline
We can finally define the set of all messages that are smaller or equal to $M$, also know as the \textit{causal history}~\cite{schwarz1994detecting}:
\begin{equation}
	\mathcal{H}(M) \defeq \{ M' : M' \leq_\mathcal{M} M \} 
\end{equation}

\subsubsection{Author-Specific Relations and Queries}
\label{sec:design:writer-rel-queries}

Given the \textit{single previous dependency} and \textit{single author} constraints on messages, it is useful to define additional relationships and queries specific to the author of messages, and therefore intended for this author's log. Again, the following relations and queries are only defined for valid messages, i.e. messages that meet the properties of Section~\ref{sec:design:message:valid}.

The happens-before relationship is similar but ignores other dependencies:
\begin{equation}
	M \anyleadsto{\textit{log}} M' \defeq \begin{cases}
		M = M'_\textit{prev} = \bot \\
		\texttt{MsgId}(M) = M'_\textit{prev}  \\
		\exists M'' : M \anyleadsto{\textit{log}} M'' \logleadsto M'
	\end{cases}
\end{equation}

The partial order and causal history for messages from the same log are analogous to those for the message graph:
\begin{equation}
\label{eq:log-message:ordering}
	M \leq_\textit{log} M' \defeq M = M' \vee M \logleadsto M'
\end{equation}
\begin{equation}
	\mathcal{H}_\textit{log}(M) \defeq \{ M' : M' \leq_\textit{log} M \} 
\end{equation}

In addition, we define a range of messages as the messages of a log that happen between two messages (including the last):
\begin{eqnarray}
	]   M, M' ]_\textit{log}~ \defeq &  \mathcal{H}_\textit{log}(M') \backslash \mathcal{H}_\textit{log}(M) 
\end{eqnarray}
This operation is similar to the commit range operation of Git (Sec.~\ref{sec:git-implementation:rel-queries}).

Since Byzantine authors may fork their logs to turn them into trees, it is useful to compute the longest prefix of both branches, i.e. the greatest lower bound of two messages:
\begin{equation}
\label{eq:log-message:log-prefix}
\begin{split}
	\texttt{LogPrefix}(M,M') \defeq & M'' : M'' \leq_\textit{log} M \wedge M'' \leq_\textit{log} M' \wedge \\
	 & \nexists M''' : M'' \logleadsto M''' \wedge M''' \leq_\textit{log} M \wedge M''' \leq_\textit{log} M'
\end{split}
\end{equation}

We can then use the first message of each branch of the fork as proofs that a fork happened:
\begin{equation}
 	\texttt{ForkProof}(M, M') \defeq \{ M'' \in ~] P, M ] ~\cup~ ] P, M' ]~: P = \texttt{LogPrefix}(M, M') \wedge P = M''_\textit{prev} \}
\end{equation}

\subsection{Log}
\label{sec:design:log}

In this section, we define a two-phase append-only log as a state-based CRDT: in the first \textit{growing} phase, a log replica is extended by appending a new message after the last one; in the second \textit{shrinking} phase, a log replica's last message is the greatest lower bound of all known forks made of valid messages. In addition, the log replica maintains a proof of the fork, in the form of a set of at least two messages that have the log's last message as predecessor. Both phases are eventually consistent: the first phase implements the same behaviour as regular append-only logs while the second  provides the longest valid prefix log among all forks known by correct replicas.

The state of a log $L$ is defined as follows: it is associated to a given author $L_\textit{author}$; it has a reference $L_\textit{last}$ to the last message from $L_\textit{author}$ that forms a strict sequence; and it has a set $L_\textit{forks}$ of messages, if any, that prove that there was a fork after $L_\textit{last}$. $L$ is in the growing phase if there are no known forks, i.e. $L_\textit{forks} = \emptyset$, otherwise it is in the shrinking phase. Once a fork as been found, $L$ stays in the shrinking phase forever. 

We first present the properties of valid logs for both phases to constrain the behaviour of Byzantine authors (Section~\ref{sec:design:log:validity}), then discuss how a log is initialized and appended to (Section~\ref{sec:design:log:appending}), then how our logs form valid state-based CRDT, with partial ordering of states (Section~\ref{sec:design:log:ordering}) and merging of different states (Section~\ref{sec:design:log:merging}). 

\subsubsection{Validity Properties}
\label{sec:design:log:validity}

The growing and shrinking phases have different validity properties. In the growing (correct) phase, in addition to the absence of fork proofs,
\begin{tcolorbox}
\begin{enumerate}
	\item[\textbf{CL1}] \textit{(no forks)}: $L_\textit{forks} = \emptyset$.
\end{enumerate}
\end{tcolorbox}

a log's last message must be valid and consistent with the log's author:
\begin{tcolorbox}
\begin{enumerate}
	\item[\textbf{CL2}] \textit{(valid last message)}: if $L_\textit{last} = M \neq \bot$ then $M$ is valid.
	\item[\textbf{CL3}] \textit{(consistent previous author)}: if $L_\textit{last} \neq \bot$ then $L_\textit{author} = (L_\textit{last})_\textit{author}$.
\end{enumerate}
\end{tcolorbox}

This protects the log from invalid message updates and makes it self-certifying (Section~\ref{sec:background:self-certification}).  Once a log has forked and is now in the shrinking phase, property \textit{CL1} is replaced by \textit{FL1}, while \textit{CL2} and \textit{CL3} still apply but have been relabeled here for clarity:
\begin{tcolorbox}
\begin{enumerate}
	\item[\textbf{FL1}] \textit{(non-empty forks)}: $L_\textit{forks} \neq \emptyset$.
	\item[\textbf{FL2}] \textit{(valid last message)}: \textit{idem CL2}
	\item[\textbf{FL3}] \textit{(consistent previous author)}: \textit{idem CL3}
\end{enumerate}
\end{tcolorbox}

In addition, similar properties now have to apply to forks: they must be valid messages, forks must have been created by $L_\textit{author}$, there must exist at least two messages with the same predecessor, and this predecessor must be the last message of the log: 
\begin{tcolorbox}
\begin{enumerate}
	\item[\textbf{FL4}] \textit{(valid forks)}: for all $M \in L_\textit{forks}$, $M$ is valid.
	\item[\textbf{FL5}] \textit{(consistent fork author)}: for all $M \in L_\textit{forks}$, $L_\textit{author} = M_\textit{author}$ and if $L_\textit{last} \neq \bot$ then $L_\textit{author} = (L_\textit{last})_\textit{author}$.
	\item[\textbf{FL6}] \textit{(valid proof)}: if $M \in L_\textit{forks}$ then $\exists M' \in L_\textit{forks}$ such that $M \neq M'$ and $M_\textit{prev} = M'_\textit{prev}$.
	\item[\textbf{FL7}] \textit{(consistent proof)}:  for all $M \in L_\textit{forks}$, $L_\textit{last} = M_\textit{prev}$. 
\end{enumerate}
\end{tcolorbox}

All other validity properties actually come from the underlying message graph (Section~\ref{sec:design:message}). Together, the validity properties of both phases constrain a Byzantine log author to only two options: either produce a correct sequential log, or produce forks of valid messages. Any other states of logs will be ignored by correct replicas.

\subsubsection{Initialization and Append}
\label{sec:design:log:appending}

The initialization and appending operation for a log are listed in Alg.~\ref{alg:log}. By convention, we use $L$ to denote the current state of a log, and $L'$ to denote the next state of log, if any. This makes the exposition and later proofs easier to read, and follows the conventions of functional programming. An implementation could update the state in-place for performance. Also by convention, for all state modifying operations, the current state of the log is the first argument to a function.

 A log is initialized from an \textit{author} and starts in the growing phase with no last message, which makes it trivially valid. A message $M$ is appended to a log $L$ only if both are valid and have consistent authors, otherwise the state of $L$ does not change. 

In the growing phase, i.e. $L_\textit{forks} = \emptyset$, when appending a new message $M$ to $L$, there are three possible cases. In the first case, when $M$ is a successor of $L_\textit{last}$, the log is updated by setting $L'_\textit{last} = M$. In the second case, when $M$ is smaller or equal to $L_\textit{last}$, i.e. it is the same or a predecessor, the message is ignored and the log is not updated. In the third case, when $M$ is concurrent with $L_\textit{last}$, then we found a new fork and the log switches to the shrinking phase.

In the shrinking phase, i.e. $L_\textit{forks} \neq \emptyset$, when appending a new message $M$ to $L$, there are two possible cases. In the first case, $M$ does not provide a new fork proof, i.e. $M$ might be a predecessor of $L_\textit{last}$ or it might belong to a branch after $L_\textit{last}$. In that case, the state of the log is not changed. Otherwise, $M$ is on a new fork that started earlier than $L_\textit{last}$. In this case, the longest common prefix (greatest lower bound) of both $L_\textit{last}$ and $M$ is computed, which is assigned to $L'_\textit{last}$ and the proof of the fork is stored in $L'_\textit{fork}$.

\begin{algorithm}
\begin{algorithmic}[1]

	\Function{Initialize$_\mathds{L}$}{$author$}
		\State $L_\textit{author} \leftarrow \textit{author}$
		\State $L_\textit{last} \leftarrow \bot$
		\State $L_\textit{forks} \leftarrow \emptyset$
	\EndFunction
	\State
	
	\Function{Append}{$L, M$}
		\State \textbf{Preconditions}:  $L$ is valid, $M$ is valid, and $M_\textit{author} = L_\textit{author}$.
		\State

		\If{$L_\textit{forks} = \emptyset$}
			\If{$L_\textit{last} \logleadsto M$} \Comment{Normal case: $M$ is newer.}
				\State $L' \leftarrow \texttt{initialize}(L_\textit{author})$
				\State $L'_\textit{last} \leftarrow M$
				\State \Return $L'$
			\ElsIf{$M \leq_\textit{log} L_\textit{last}$} \Comment{$M$ is the same or a predecessor.}
				\State \Return $L$	
			\EndIf
		\ElsIf{$L_\textit{forks} \neq \emptyset \wedge M \nparallel_\textit{log} L_\textit{last}$} \Comment{$M$ does not provide a new fork proof.}
			\State \Return $L$
		\EndIf
		\State
		\State $L' \leftarrow \texttt{initialize}(L_\textit{author})$ \Comment{Forked cases: $L_\textit{forks} \neq \emptyset$ or $L_\textit{last} ~\parallel_\textit{log}~ M$} \label{alg:log:fork-handling}
		\State $L'_\textit{last} \leftarrow \texttt{LogPrefix}(L_\textit{last},M)$ 
		\State $L'_\textit{forks} \leftarrow \texttt{ForkProof}(L_\textit{last},M)$	
		\State \Return $L'$
	\EndFunction
\end{algorithmic}
\caption{\label{alg:log} Log: State and Operations}
\end{algorithm}

\subsubsection{Ordering}
\label{sec:design:log:ordering}

\begin{algorithm}
\begin{algorithmic}[1]
\Function{$\leq_{\mathds{L}}$}{$L$, $L'$}
	\State \textbf{Preconditions}: $L$ and $L'$ are valid, and $L_\textit{author} = L'_\textit{author}$.
	\State
	\If{$L_\textit{forks} = \emptyset \wedge L'_\textit{forks} = \emptyset$}
	
		\State \Return $L_\textit{last} \leq_\textit{log} L'_\textit{last}$ \Comment{See Alg.~\ref{alg:log}}
	\ElsIf{$L_\textit{forks} = \emptyset \wedge L'_\textit{forks} \neq \emptyset$}
		\State \Return \textbf{true}  \Comment{A forked log state is always larger than a non-forked state.}
	\ElsIf{$L_\textit{forks} \neq \emptyset \wedge L'_\textit{forks} = \emptyset$}
		\State \Return \textbf{false} \Comment{$L'$ is still in growing phase but not $L$, therefore cannot be greater}
	\Else
		\State \Return $L'_\textit{last} \leq_\textit{log} L_\textit{last}$ \Comment{Is $L'$ the same length as, or shrunk more than $L$?}
	\EndIf
\EndFunction	
\end{algorithmic}
\caption{\label{alg:log:ordering} Log: Ordering}
\end{algorithm}

We now define a partial order over all possible log states, listed in Alg.~\ref{alg:log:ordering}, as the first step in establishing logs as state-based CRDTs (Section~\ref{sec:background:crdt}). As for all other operations, the less or equal relationship between two log states, $L \leq_\mathds{L} L'$, is only defined for valid logs and the two states must have consistent authors. There are four cases to consider:
\begin{enumerate}
	\item Both $L$ and $L'$ are in the growing phase. In this case, $L$ is smaller or equal to $L'$ if and only if $L_\textit{last}$ is equal or a predecessor of $L'_\textit{last}$, i.e. $L_\textit{last} \leq_\textit{log} L'_\textit{last}$ (Eq.~\ref{eq:log-message:ordering}).

	\item $L$ is in the growing phase and $L'$ is in the shrinking phase. Regardless of their states, then $L$ is smaller than $L'$.

	\item $L$ is in the shrinking phase and $L$ is in the growing phase. This is the opposite of the second case, therefore $L$ is not smaller or equal to $L'$.
	\item Both $L$ and $L'$ are in the shrinking phase. This is the opposite of the first case, therefore $L$ is smaller or equal if $L'_\textit{last}$ is smaller or equal to $L_\textit{last}$. 
\end{enumerate}

A structured proof that is $\leq_\mathds{L}$ is a partial order is given Section~\ref{sec:proofs:convergence:log}. As previously defined, $\leq_\mathds{L}$ is a not a total order on all possible log states for two reasons. First, it is only defined on log states from the same author and log states from different authors are incomparable. Second, when the last message of both states are on different branches, i.e. $L_\textit{last} \parallel L'_\textit{last}$, neither of the states is smaller than the other.

There is an upper bound on all possible states that corresponds to a fork on the first message of the log,  i.e. $L_\textit{last} = \bot$ and for all $M \in L_\textit{forks}$, $M_\textit{prev} = \bot$. However, there is an infinite number of intermediate states between the initial state and the upper bound that corresponds to an infinite number of correct logs with arbitrarily large number of messages, prior to shrinking.

\subsubsection{Merging}
\label{sec:design:log:merging}

The next step in our design is a function $\sqcup_\mathds{L}$, listed in Alg~\ref{alg:log:merging}, to merge any two states of logs $L,L'$ to obtain a new state $L''$ that includes updates from both. This may be used, for example, to reconcile the state of a log within a replica with that of another replica.  Similar to other operations, it is only defined for valid log states with consistent authors. 

\begin{algorithm}[h]
\begin{algorithmic}[1]
\Function{$\sqcup_{\mathds{L}}$}{$L$, $L'$}
	\State \textbf{Preconditions}:  $L_\textit{author} = L'_\textit{author}$, $L$ and $L'$ are valid.
	\State
	\If{$L_\textit{last} ~\nparallel_\textit{log}~ L'_\textit{last}$}
		\If {$L_\textit{forks} = \emptyset \wedge L'_\textit{forks} = \emptyset$} \Comment{No known forks}
			\If{$L_\textit{last} \leq_\textit{log} L'_\textit{last}$}
				 \State \Return $L'$
			\Else
				 \State \Return $L$
			\EndIf
		\ElsIf{$L_\textit{forks} \neq \emptyset \wedge L'_\textit{forks} = \emptyset$}
			 \State \Return L 
		\ElsIf{$L_\textit{forks} = \emptyset \wedge L'_\textit{forks} \neq \emptyset$}
			 \State \Return $L'$ 
		\EndIf
	\EndIf
	\State
	\Comment{Either $L_\textit{last} \parallel_\textit{log} L'_\textit{last}$ or both are forked.}
	\State $L'' \leftarrow \texttt{initialize}_\mathds{L}(L_\textit{author})$
	\State $L''_\textit{last} \leftarrow \texttt{LogPrefix}(L_\textit{last},L'_\textit{last})$
	\State $L''_\textit{forks} \leftarrow \{ M \in (\texttt{ForkProof}(L_\textit{last},L'_\textit{last}) \cup L_\textit{forks} \cup L'_\textit{forks}) : M_\textit{prev} = L''_\textit{last} \}$
	\State \Return $L''$	
\EndFunction
\end{algorithmic}
\caption{\label{alg:log:merging} Log: Merging}
\end{algorithm}

There are multiple cases to consider depending on whether $L$ and $L'$ are growing or shrinking, and whether $L_\textit{last}$ and $L'_\textit{last}$ are on different branches of forks, i.e. $L_\textit{last} \parallel_\textit{log} L'_\textit{last}$.  If both $L$ and $L'$ are already forked or  $L_\textit{last} \parallel_\textit{log} L'_\textit{last}$, then the fork resolution applies, similar to the case of a fork between $L_\textit{last}$ and a message $M$ in \textsc{Append} (Section~\ref{sec:design:log:appending}): $L''_\textit{last}$ is the longest prefix (greatest lower bound) of both $L_\textit{last}$ and $L'_\textit{last}$, and $L''_\textit{forks}$ contains either a new fork proof or a superset for fork proofs that apply. 

Otherwise, $L''$ is equal to the greatest state of $L$ or $L'$. There are three possible cases (the fourth case was already covered by the previous paragraph):
\begin{enumerate}
	\item Both $L$ and $L'$ are still growing. Since $L_\textit{last} \nparallel_\textit{log} L'_\textit{last}$, then either $L$ is greater or equal to $L'$, in which case $L''=L$, or $L'$ is greater than $L$ in which case $L'' = L'$.
	\item $L$ is shrinking but $L'$ is still growing. Then $L''=L$.
	\item $L$ is growing but $L'$ is shrinking. Then $L''=L'$.
\end{enumerate}

To complete this second step, we need to show that $L''=L \sqcup_\mathds{L} L'$ computes the least upper bound of $L$ and $L'$ according to our partial order ($\leq_\mathds{L}$, Section~\ref{sec:design:log:ordering}). Informally, this is the case because either $L''$ is equal to the greatest of $L$ or $L'$ which is trivially a least upper bound. For all other cases that result in $L''$ being in a shrinking phase, $L''_\textit{last}$ will be the longest prefix of both $L_\textit{last}$ and $L'_\textit{last}$ (by definition of \texttt{LogPrefix} in Eq.~\ref{eq:log-message:log-prefix}), itself a greatest lower bound, that result in $L''$ being the least upper bound. A more detailed proof is given in Section~\ref{sec:proofs:convergence:log}.

\subsubsection{Monotonicity}

The last step in establishing our log as a state-based CRDT is to show that all operations are \textit{monotonic}, i.e. they result in a state that is equal or larger than the input states. This is trivially the case for the merge operation, by definition of a least upper bound and otherwise there is only one other state-changing operation, which is \textsc{Append}. This is easy to verify, since either the input state is returned, which is equal, or a new state $L'$ that is larger is returned, either because $L'$ is still growing and $L_\textit{last}$ is larger, or $L'$ is shrinking and $L'_\textit{last}$ is smaller. A more detailed proof is given in Section~\ref{sec:proofs:convergence:log}. Given this completed last step, our log design satisfies all requirements of a state-based CRDT. And because state-based CRDTs are tolerant to Byzantine processes (Section~\ref{sec:background:byzantine-state-crdt}), our log design is a Byzantine fault-tolerant state-based CRDT. 

\subsection{Frontier: Set of Logs}
\label{sec:design:frontier}

We now show how to track the latest state of a set of authors with a \textit{frontier} state-based CRDT. A frontier is essentially a grow-only set of logs, with the constraint that at most one log state, the largest received so far, is maintained per author. It behaves similarly as a grow-only dictionary of CRDTs (such as \cite{lavoie2023inftypset, lavoie2023gocledger}), but we prefer the set formulation to avoid the redundant use of \textit{authors} as keys, that are already part of the log state.

\subsubsection{Validity Properties}

Frontiers are simple and straight-forward sets, so the only two validity requirements are that logs stored are valid and there is at most one log per author:
\begin{tcolorbox}
\begin{enumerate}
	\item[\textbf{F1}] \textit{(valid logs)}: for all $L \in F$, $L$ is valid.
	\item[\textbf{F2}] \textit{(at most one log per author)}: for all $L \in F$, $\nexists L' \in F : L' \neq L \wedge L_\textit{author} = L'_\textit{author}$.
\end{enumerate}
\end{tcolorbox}

\subsubsection{State and Operations}

The operations on a frontier are listed in Alg.~\ref{alg:frontier}. By convention, for all state modifying operations, the current state $F$ is the first argument of the function and a different value is returned. As for logs, an implementation may choose to mutate a frontier in place for better performance.

A frontier is initialized as an empty set. The full message history of a frontier is queried with $\texttt{Messages}(F)$, which consists of the union, for all $L$ in $F$, of the causal history of $L_\textit{last}$ and all messages in $L_\textit{forks}$. The state of a log $L$ within a frontier $F$ is updated with $\textsc{Update}(F,L)$: if $L_\textit{author}$ is already present in some log $L'$ in $F$, the new state is the merged state of $L$ and $L'$, i.e. $L \sqcup_\mathds{L} L'$, otherwise $L$ is simply added to $F$. Similar to grow-only dictionaries of CRDTs~\cite{lavoie2023inftypset, lavoie2023gocledger}), $F$ is smaller or equal to $F'$, i.e.  $F \leq_\mathds{F} F'$, if and only if $F'$ has a superset of the authors of $F$ and the state of each log associated to each author in $F$ is smaller or equal to the log associated to the same author in $F'$. Finally, two frontiers are merged by merging the state of the corresponding logs for authors that are present in both, and adding the logs for authors that are present in only one of $F$ or $F'$.

The proof that a frontier is a state-based CRDT is given in Section~\ref{sec:proofs:convergence:frontier} and the proof that all frontier operations maintain the validity properties is given in Section~\ref{sec:proofs:valid-frontier}. They are straightforward and require no additional exposition so we simply refer the reader to them.

\begin{algorithm}
\begin{algorithmic}[1]
	\Function{Initialize$_\mathds{F}$}{}
		\State \Return $\emptyset$ \Comment{Set of local logs}
	\EndFunction
	\State

	\Function{Messages}{$F$} \Comment{Set of messages reachable from $F$}
		\State \textbf{Preconditions}: $F$ is valid.
		\State
		\State \Return $\bigcup\limits_{L \in F} (\mathcal{H}(L_\textit{last}) \cup \bigcup\limits_{M \in L_\textit{forks}} \mathcal{H}(M))$ 
	\EndFunction
	\State	
	
	\Function{Update}{$F, L$}
		\State \textbf{Preconditions}: $F$ and $L$ are valid.
		\State
		\If{$\exists L' \in F : L'_\textit{author} = L_\textit{author}$}
			\State \Return $F \backslash \{ L' \} \cup \{ L \sqcup_\mathds{L} L' \}$
		\Else
			\State \Return $F \cup \{ L \}$
		\EndIf
	\EndFunction
	\State
	
	\Function{$\leq_\mathds{F}$}{$F,F'$}
		\State \textbf{Preconditions}: $F$ and $L$ are valid.
		\State
		\State $A \leftarrow \{ L_\textit{author} : L \in F \}$
		\State $A' \leftarrow \{ L_\textit{author} : L \in F' \}$
		\State $\mathcal{L} \leftarrow \{ (L, L') : L \in F \wedge L' \in F' \wedge L_\textit{author} = L'_\textit{author} \wedge L_\textit{author} \in A \}$
		\State \Return $A \subseteq A' \wedge  \bigwedge\limits_{(L,L') \in \mathcal{L}} L \leq_\mathds{L} L'$
	\EndFunction
	\State
	
	\Function{$\sqcup_\mathds{F}$}{$F,F'$}
		\State \textbf{Preconditions}: $F$ and $F'$ are valid.
		\State
		\State $\mathcal{L} \leftarrow \{ L : L \in F \wedge \nexists L' \in F' : L_\textit{author} = L'_\textit{author} \}$
		\State $\mathcal{L}' \leftarrow \{ L' : L' \in F' \wedge \nexists L \in F : L_\textit{author} = L'_\textit{author} \}$
		\State $\mathcal{L}'' \leftarrow \{ L \sqcup_\mathds{L} L' : L \in F \wedge L' \in F' \wedge L_\textit{author} = L'_\textit{author} \}$
		
		\State \Return $\mathcal{L} \cup \mathcal{L'} \cup \mathcal{L}''$
	\EndFunction
\end{algorithmic}
\caption{\label{alg:frontier} Frontier: State and Operations}
\end{algorithm}

\subsubsection{Impact of Forked Logs}

A forked log looses liveness with every log replica that has a proof it has forked: messages appended to any of the branches are ignored by correct replicas. There is however a window of opportunity between the detection of a fork by the first replica and the propagation of the fork proof to all replicas. During that time window, a malicious replica may still extend the branches that are replicated by unaware replicas. Nonetheless, even if those branches are extended and replicated, they will only influence the state of correct (non-forked) replicas if an explicit dependency is recorded in correct logs towards malicious branches. Once the fork proofs have fully replicated, the only other option a malicious participant has is to reveal or introduce new forks earlier in their log. But again, these additional forks can only influence the state of correct logs if dependencies are recorded. We therefore see that although not negligible, the impact an adversary can have by forking their log is limited both in time and in the explicit dependencies that correct logs record.

Also, because the dependencies of messages on correct (growing) logs may include branches of forked logs, the branches after the last message of forked logs still need to be replicated to properly construct the causal history of messages from correct logs, as performed with $\texttt{Messages}(F)$. This is a necessary property, in an accounting application~\cite{lavoie2023gocledger} for example, to compute how many tokens were double-spent in forked branches and be able to properly repair the damage.

To ensure a malicious author cannot continue to corrupt the state of correct logs after a fork is discovered, a correct author can simply stop listing as dependencies the messages from authors that have forked their log. This way, even if a malicious author were to intentionally produce earlier forks, only the proofs will be replicated and none of the other newer messages is be depended and can therefore be safely ignored by correct replicas.

\newpage
\section{Implementation over Git}
\label{sec:git-implementation}

In this section, we explain at a high-level how to implement our design over Git~\cite{git}. This is straight-forward since Git provides a data model with commit histories and low-level commands necessary to easily implement our algorithms. The full prototype in still a work in progress, and this paper will be revised once it is complete, but we expect it may take at most a few hundred lines of Bash code and should therefore be easily portable to other languages that have bindings to libgit2~\cite{libgit2}.

\subsection{Messages as Git Commits}

We first map our messages (Table~\ref{tb:message-schema})  to Git commits. Git commits have the following attributes~\cite{git-commit}:
\begin{itemize}
	\item \textit{tree-hash}: reference to an immutable tree that represents the state of the filesystem at the time of the commit. 
	\item \textit{parents}: reference to other commits on which this commit depends;
	\item \textit{author}: \textit{name} and \textit{email} address of the person who created the content of the commit (in the case of the linux kernel, someone who submitted a patch, potentially by email to one of the maintainers);
	\item \textit{committer}: \textit{name} and \textit{email} address of the person who created the commit and added it to the repository;
	\item \textit{message}: description of the commit content.
\end{itemize}

The mapping is straight-forward and shown in Table~\ref{tb:git-commits}: the only significant difference being that $M_\textit{prev}$ and $M_\textit{deps}$ are combined together within the \textit{parents} field.

\begin{table}[ht]
\centering
\caption{Encoding of messages as Git commits.}
\label{tb:git-commits}
\begin{centering}
\begin{tabular}{lll}
\textbf{Git Commit Field}         & \textbf{Purpose} \\ \hline
\textit{author} 		      	      &  $M_\textit{author}$ as \textit{author.name}, the \textit{author.email} is not used;      \\
\textit{committer}                     & Not used;           \\
\textit{parents}    	   	      & First parent is always $M_\textit{prev}$ while the next are $M_\textit{deps}$;           \\
\textit{message}  	  	      &  Contains both the $M_\textit{payload}$ and $M_\textit{signature}$; \\
\textit{tree-hash}  		      & Not used, left open for applications.                             
\end{tabular}
\newline
\end{centering}
\end{table}

Because commits are signed, it is therefore not possible for an adversary to forge alternative commits for authors for which they do not possess the private key.

\subsection{Log's Last Message as a Self-Certifying Branch Reference}

Git references are essentially pointers to commits~\cite{git-reference}. In the case of branches, those references are mutable so that the same name will point to different commits over time. Because a log has a constant \textit{author}, we use the author's public key in a branch name,  \texttt{<author>/last}, to reference $L_\textit{last}$, the latest \textit{sequential} (non-forked) message  from that author (Section~\ref{sec:design:log}). An additional benefit of using the author's public key as a branch name, is that the branch is now self-certifying (Section~\ref{sec:background:self-certification}): when another replica provides the latest state of a log under a branch name, the receiver can check that this branch name is consistent with the author of the commits pointed at, which themselves are signed by the author. 

An adversarial replica can only provide existing alternative commits from the same correct author, as alternative referents to the branch name, because they do not have access to the correct author's private key to sign messages.They can therefore hamper liveness, by not supplying the latest commits and therefore delaying their propagation, but they cannot wrongly attribute commits not signed with the \textit{author}'s private key as an alternative log history.

\subsection{Fork Proofs as Fork Commit and a Self-Certifying Branch}

Similar to the encoding of the last message as a self-certifying branch, we encode $L_\textit{forks}$, i.e. the set of fork proofs, in a branch. To represent the set we use a \textit{fork commit} (Table~\ref{tb:git-fork-commit}) whose parents point to the first commit of each branch of the fork, themselves having the commit referenced by  \texttt{<author>/last} as the same parent. This fork commit therefore forms a "diamond" with $L_\textit{last}$.

We then make the fork proof self-certifying by storing it under the branch \texttt{<author>/forks/<last-id>} in which \texttt{<last-id>} is the commit id  referenced by \texttt{<author>/last}.  This way we can quickly test whether a log is in a forked state by dereferencing \texttt{<author>/last} to obtain a \texttt{<last-id>} and then checking whether the branch \texttt{<author>/forks/<last-id>} exists and points to a valid fork commit. A fork commit is valid if it has at least two parents that are valid commits and that both of them have \texttt{<last-id>} as parent.

\begin{table}[ht]
\caption{Encoding of fork proofs as Git commits.}
\label{tb:git-fork-commit}
\begin{centering}
\begin{tabular}{lll}
\textbf{Git Commit Field}         & \textbf{Purpose} \\ \hline
\textit{author} 		      	      &  $M_\textit{author}$ as \textit{author.name}, the \textit{author.email} is not used.      \\
\textit{committer}       	      & Not used. \\
\textit{parents}    	   	      & \begin{tabular}{@{}l@{}}Each parent references the first commit of a branch that itself has \\ \texttt{<author>/last} as parent. \end{tabular} \\
\textit{message}  	  	      & "Fork proof" message type (signature not necessary). \\
\textit{tree-hash}  		      & Not used, left open for applications.                             
\end{tabular}
\newline
\end{centering}
\end{table}


A fork commit does not need to be signed since only the existence the two parents commits from the author of the fork matters. Moreover, in practice it may not be necessary to keep all fork proofs, any two messages that form a valid proof suffice, so a correct replica may simply keep the first valid fork commit it either created or replicated. This avoids mutating branch references to a different commit, which needs a forced update when using the default replication behaviour of Git. 

\subsection{Handling Correctly Signed but Invalid Messages}

The implementation has to deal with an issue that does not arise in our previous presentation because we assume the algorithms are only defined for valid messages, logs, and frontiers: an author can correctly sign a message that is invalid. This results in neither a correct, growing, log, nor a forked log. 

For simplicity, we ignore invalid messages keeping \texttt{<author>/last} to the last valid message, and allow the author to sign a new valid message that would replace the invalid one. However, the production of an invalid message that is correctly signed is still a proof of misbehaviour from the author, so we store this proof under \texttt{<author>/invalid}. A correct replica keeps at most one such invalid message per author to bound the amount of storage Byzantine authors may require from correct replicas. This invalid proof is replicated between correct replicas, which may choose to stop replicating authors of invalid messages.

\subsection{Tracking Remote Replicas with Remote Branches}

Similar to existing Git conventions, we use the \texttt{remotes/$<$origin$>$} prefix to track remote branches and efficiently compute missing commits between two frontiers (which is equivalent to computing $\texttt{Messages}(F) \backslash \texttt{Messages}(F')$ using Alg.~\ref{alg:frontier}). For hosted replicas accessed over secure urls, either using the Git protocol or https, \texttt{<origin>} is the local label for that url, as typically added with \texttt{git remote add origin <origin> <url>}. A peer-to-peer alternative is also possible: \texttt{<origin>} may also be the public key of the replica, which may or may not be the same as one of the logs, and authenticated with a protocol such as Secret-Handshake~\cite{tarr2015shs}. While the latter is not commonly used among developers, Git has a well-defined extension mechanism to define such protocols~\cite{git-protocols}, so this extension should be relatively straight-forward to implement.

\subsection{Validating Messages and Forks Incrementally}

Ideally, the replication protocols would validate the commits as they are received and would discard those that do not have the expected properties (Section~\ref{sec:design:message}). In that case, updating the \texttt{remotes/<origin>} branches would be sufficient. This however requires a custom replication protocol.

For the current implementation, instead, we separate the validation from the replication process to ease implementation and reuse the existing Git replication protocols. The commits under \texttt{remotes/<origin>} are therefore not trusted until validated and a separate prefix, \texttt{valids/<origin>}, is used to track the commits that are valid. Only valid logs are used for ordering and merging operations (Alg.~\ref{alg:log:ordering} and ~\ref{alg:log:merging}).

While requiring less additional custom machinery this approach has the drawback that invalid commits are stored before being validated, so this opens an attack vector in which Byzantine replicas may produce an arbitrary large number of invalid commits during replication. A correct replica would have to limit the size of updates it accepts to limit the damage a Byzantine replica may do. We leave a full hardening of the protocol against resource exhaustion for future work.

While not necessary to ensure convergence, as well as safety and liveness properties, our implementation enforces an additional constraint on messages by ensuring that every dependency from the same author is equal or larger for later messages:

\begin{tcolorbox}
\begin{enumerate}
	\item [\textbf{M7}] \textit{(monotonic dependencies)}: if $M \logleadsto M'$ then for all $(\textit{msgId}, \textit{msgId}') \in M_\textit{deps} \times M'_\textit{deps} : \textit{msgId} = \texttt{MsgId}(D) \wedge \textit{msgId}' = \texttt{MsgId}(D') \wedge D_\textit{author} = D'_\textit{author} \Rightarrow D \leq_\mathcal{M} D'$.
\end{enumerate}
\end{tcolorbox}

This is the natural structure of applications built using append-only logs and those that follow \textit{causal history} (Section~\ref{sec:design:message}).

\subsection{Frontier as a Set of Branches}

A Frontier (Alg.~\ref{alg:frontier}) is therefore simply a set of branches representing logs and the commits they point to. We store the logs of a replica \texttt{locals/<author>/*} in contrast to the \texttt{remotes/<origin>/<author>/*} and \texttt{valids/<origin>/<author>/*} prefixes. The distinction between \texttt{locals/<author>/*} and \texttt{valids/<origin>/*} is necessary because two different origins may replicate logs that are valid and non-forked individually but actually represent different branches, so the log under \texttt{locals/<author>} will be forked (shrinking) instead. A replica therefore maintains one frontier of not yet validated logs and one frontier of validated logs for each remote origin, as well as a single local frontier combined from the validated logs of all origins.

\subsection{Relationships and Queries}
\label{sec:git-implementation:rel-queries}

In Git, the most commonly used elementary query, \texttt{git log}, is actually the causal history of a commit (which is equivalent to $\mathcal{H}(M)$ from Section~\ref{sec:design:queries}). We therefore define the other relationships and queries in terms of this. Git also provides a convenient \texttt{git merge-base} command that covers most of the rest of our needs. All key relationships and queries we use in the previous algorithms, as well as their Git equivalent, are summarized in Table~\ref{tb:git-rel-queries}.

\begin{table}[h]
\begin{centering}
\caption{Git relationships and queries. Assume $\texttt{<c-id>} = \texttt{MsgId}(M)$, $\texttt{<c-id-2>} = \texttt{MsgId}(M')$, and $M_\textit{author} = M'_\textit{author} = \texttt{<author>}$. Bash commands representing binary relations return $0$ if true, $>0$ otherwise. See documentation for various formats with which to return the commit sets.  Tested with git version 2.24.3, commands and syntax might change for other git versions. \newline}
\label{tb:git-rel-queries}
\renewcommand{\arraystretch}{1.3}
\begin{tabular}{lll}
\textbf{Relationships}         & \textbf{Git+Bash/Unix Equivalent} \\ \hline \hline

$M \leq_\mathcal{M} M'$            & \texttt{git merge-base --is-ancestor <c-id> <c-id-2>} \\ \hline

$M \leadsto M'$          	          & \begin{tabular}{@{}l@{}}
						     \texttt{git merge-base --is-ancestor <c-id> <c-id-2> \&\&} \\
						     \texttt{{[} "<c-id>" != "<c-id-2>" {]} }  						     \end{tabular} \\ \hline
				     
$M \leq_\textit{log} M'$                & \begin{tabular}{@{}l@{}}
						     \texttt{git log <c-id-2> --author=<author> --format=\%H |} \\
						     \texttt{grep -q <c-id>} 
						      \end{tabular} \\ \hline
						      
$M \logleadsto M'$          	          & \begin{tabular}{@{}l@{}}
						   \texttt{git log <c-id-2>\textasciitilde1 --author=<author> --format=\%H |} \\
						   \texttt{grep -q <c-id>}  
						   \end{tabular} \\ \hline
						     
~\\

\textbf{Queries}         & \textbf{Git+Bash/Unix Equivalent} \\ \hline \hline

$\mathcal{H}(M)$    		          & \texttt{git log <c-id>} \\ \hline
						   
$\mathcal{H}_\textit{log}(M)$      & \begin{tabular}{@{}l@{}}
						     \texttt{git log <c-id> --author=<author> --format=\%H }
						     \end{tabular} \\ \hline
						     
 ${]}M,M'{]}_\textit{log}$            	  &  \texttt{git log <c-id>..<c-id-2> --author=<author>}  \\ \hline
						     
$\texttt{LogPrefix}(M,M')$             & \texttt{git merge-base <c-id> <c-id-2>} \\ \hline

$\texttt{ForkProof}(M,M')$            &  \begin{tabular}{@{}l@{}}
							\texttt{P=\$(git merge-base <c-id> <c-id-2>);} \\
							\texttt{git rev-list --children --all --author=<author> \$P | }\\
							\texttt{grep -v \$B}	
							\end{tabular} \\ \hline	                               
\end{tabular}
\newline
\end{centering}
\end{table}

We believe implementing our algorithms with the previous branch conventions and equivalent git commands should pose no special conceptual difficulties, so this completes our presentation of the design. Implementers should be able to focus on other important software engineering requirements (performance, maintainability, documentation, portability, etc.) which we glossed over. We will revise this document if actual practice teaches us we overlooked some important aspects.

\newpage
\section{Proofs}
\label{sec:proofs}

In this section, we provide detailed proofs for convergence, safety, and liveness of our design. They are written with significant detail because this approach helped us find previously ignored corner cases, therefore serving a similar but more systematic purpose as unit testing.

\subsection{Definitions}

We first make the semantics of algorithms more precise with the following definitions:
\begin{proof}
	\begin{pfenum}
		\item $\mathds{A}$ is the set of all possible authors.
		\item $\mathds{M}$ is the set of all possible valid messages.
		\item $\mathds{L}$ is the set of all possible valid log states.
		\item $\mathds{L}(\textit{author}) : \textit{author} \in \mathds{A}$ is the set of all possible valid log states from \textit{author}, i.e. $\mathds{L}(\textit{author}) \subseteq \mathds{L}$ such that $L,L' \in \mathds{L}(\textit{author}) \Rightarrow L_\textit{author} = L'_\textit{author} = \textit{author}$.
		\item $\mathds{F}$ is the set of all possible valid frontiers.
	\end{pfenum}
\end{proof}

\subsection{Convergence}

To establish the convergence of both logs and frontiers, we need to establish that their states form a \textit{monotonic semi-lattice}~\cite{Shapiro2011CRDTs}, which involves three propositions: First, that all possible states can be partially ordered by $\leq$. This is a requisite for the next two properties. Second, that merging two states computes their \textit{Least Upper Bound} (LUB) in that partial order. This ensures that the merge is \textit{commutative}, \textit{associative}, and \textit{idempotent}, providing \textit{safety}, \textit{i.e.}, that replicas will agree on the final state regardless of ordering, delays, or duplication of merge operations. Third, that all operations modify the state $S$ of a replica such that the new state $S'$ is either equal or larger than the previous state $S$ in the partial order (\textit{monotonicity}). This ensures all state changes will be eventually reflected in the new state of all replicas, either because the same update(s) will have concurrently been applied or because the new state will be the result of a merge. Assuming an underlying communication medium that ensures new states to be eventually delivered to other replicas, the three propositions combined ensure both \textit{liveness} and \textit{safety}: all state changes are going to be replicated on all replicas \textit{and} all replicas will agree on the final state automatically, i.e. \textit{strong eventual consistency}~\cite{Shapiro2011CRDTs}.

\subsubsection{Messages}
\label{sec:proofs:messages}

\prove{\textbf{Ordering}: $\leq_\textit{log}$ (Eq.~\ref{eq:log-message:ordering}) partially orders $\mathds{M}$.}
\begin{proof}
	\step{}{\textbf{Reflexivity}:
		\assume{$M \in \mathds{M}$}
		\prove{$M \leq_\textit{log} M$ is always true.}
	}
	\begin{proof}
		By the definition of $\leq_\textit{log}$ which includes $M = M'$.
	\end{proof}
	
	\step{}{\textbf{Transitivity}: 
		\assume{\begin{pfenum}
			\item $M,M',M'' \in \mathds{M}$
			\item $M \leq_\textit{log} M'$
			\item $M' \leq_\textit{log} M''$
		\end{pfenum}}
		\prove{$M \leq_\textit{log} M''$}}
	\begin{proof}
		\case{$M = M' = M''$}
		\begin{proof}
			$M = M''$, which makes the first condition on $M \leq_\textit{log} M''$ true.
		\end{proof}
	
		\case{$M = M' \wedge M' \leq_\textit{log} M''$}
		\begin{proof}
			By substituting $M'$ with $M$, therefore $M \leadsto M''$ which makes the second condition on $M \leq_\textit{log} M''$ true. 
		\end{proof}
		
		\case{$M \logleadsto M' \wedge M' = M''$}
		\begin{proof}
			By substituting $M'$ with $M''$, therefore $M \leadsto M''$ which makes the second condition on $M \leq_\textit{log} M''$ true. 
		\end{proof}
		
		\case{$M \logleadsto M' \wedge M' \logleadsto M''$}
		\begin{proof}
			This is the third (transitive) condition on $M \logleadsto M''$, therefore $M \leadsto M''$ which makes the second condition on $M \leq_\textit{log} M''$ true. 
		\end{proof}
		
		\qedstep{}
		\begin{proof}
			True for all possible cases, therefore always true.
		\end{proof}
		
	\end{proof}
	
	\step{}{\textbf{Antisymmetry}:
		\assume{\begin{pfenum}
			\item $M, M' \in \mathds{M}$
			\item $M \leq_\textit{log} M'$
			\item $M' \leq_\textit{log} M$
		\end{pfenum}}
		\prove{$M = M'$}
	}
	\begin{proof}
		There are two alternative conditions for $\leq_\textit{log}$ to be true: either both are equal or one $\logleadsto$ the other. If they are not equal, then there would be a cycle between $M$ and $M'$ which would violate the acyclic validity condition that would contradict the fact that $M,M' \in \mathds{M}$ implies they are valid. Therefore the only possibility left is that they are equal.
	\end{proof}
	
	\qedstep{}
	\begin{proof}
		The three properties are the definition of a partial order.
	\end{proof}
\end{proof}

\subsubsection{Log}
\label{sec:proofs:convergence:log}

\prove{For any given \textit{author}, the Log design listed in Alg.~\ref{alg:log}, ~\ref{alg:log:ordering}, and \ref{alg:log:merging} is a state-based (convergent) CRDT.}

\begin{proof}
	\define{\begin{pfenum}
		\item $L,L' \in \mathds{L}(\textit{author})$
		\item $M \in \mathds{M}$
	\end{pfenum}}
		
	\step{}{\textbf{Ordering}:  $\leq_\mathds{L}$ (Alg.~\ref{alg:log:ordering}) partially orders $\mathds{L}(\textit{author})$.}
	\begin{proof}

		\pfsketch~  By cases on the different phases of the logs.

		\step{}{\case{$L_\textit{forks} = L'_\textit{forks} = \emptyset$}}
		\begin{proof}
			$L$ and $L'$ are only distinguished by their last messages, respectively $L_\textit{last}$ and $L'_\textit{last}$, that are partially ordered by $\leq_\textit{log}$ (Section~\ref{sec:proofs:messages}). 
		\end{proof}
		
		\step{}{\case{$L_\textit{forks} = \emptyset$ and $L'_\textit{forks} \neq \emptyset$}}
		\begin{proof}
			$L'$ is a (valid) forked log, which corresponds to the second possible phase of a log, while $L$ is still in the first phase. $L'$ is always greater than $L$ which forms a total order between any possible states of $L$ and $L'$.
		\end{proof}
		
		\step{}{\case{$L_\textit{forks} \neq \emptyset$ and $L'_\textit{forks} = \emptyset$}}
		\begin{proof}
			Opposite of the previous case, which also forms a total order between any possible states of $L$ and $L'$.
		\end{proof}
		
		\step{}{\case{$L_\textit{forks} \neq \emptyset$ and $L'_\textit{forks} \neq \emptyset$}}
		\begin{proof}
			Similar to the first case with reversed $L_\textit{last}$ and $L'_\textit{last}$, i.e. a forked log is considered larger or equal if there exists a proof that it forked at the same or an earlier point on the same message sequence. 
		\end{proof}
		
		\qedstep{}
		\begin{proof}
			In all cases, all possible valid states of $L$ and $L'$ are partially or totally-ordered therefore $\leq_\mathds{L}$ defines a partial order over  $\mathds{L}$.
		\end{proof}
	\end{proof}
	
	\step{}{\textbf{Least-Upper Bound}:	
		\prove{$L'' = L \sqcup_\mathds{L} L'$ is the LUB of $L$ and $L'$ in $\mathds{L}(\textit{author})$ partially ordered by $\leq_\mathds{L}$.}
	}
	\begin{proof}
		\step{}{\case{$L_\textit{forks} = L'_\textit{forks} = \emptyset$}}
		\begin{proof}
			\step{}{\case{$L_\textit{last} \leq_\textit{log} L'_\textit{last} \vee L'_\textit{last} \leq_\textit{log} L_\textit{last} $ (equivalent to $L_\textit{last} \nparallel_\textit{log} L'_\textit{last}$)}}
			\begin{proof}			
				\step{}{\case{$L_\textit{last} \logleadsto L'_\textit{last}$}}
				\begin{proof}
					By definition both $L$ and $L'$ are valid and non-forked, therefore the first case of $\sqcup_\mathds{L}$ applies and \texttt{Append} is called with a valid $M=L'_\textit{last}$ and returns $L''$. $L''_\textit{last} = L'_\textit{last} \geq_\textit{log} L_\textit{last}$ and $L''_\textit{forks} = \emptyset$. By definition, $L \leq_\mathds{L} L' = L''$ and is a least upper bound of both $L$ and $L'$.
				\end{proof}
			
				\step{}{\case{$L'_\textit{last} \leq_\textit{log} L_\textit{last}$}}
				\begin{proof}
					By definition both $L$ and $L'$ are valid and non-forked and, therefore the first case of $\sqcup_\mathds{L}$ applies and \texttt{Append} is called with a valid $M=L_\textit{last}$ and returns $L''$. $L''_\textit{last} = L_\textit{last} \geq_\textit{log} L'_\textit{last}$ and $L''_\textit{forks} = \emptyset$. By definition, $L' \leq_\mathds{L} L = L''$ and is a least upper bound of both $L$ and $L'$.
				\end{proof}
				
				\qedstep
				\begin{proof}
					All possible cases induced by the definition of $\leq_\textit{log}$ are covered: The case where $L_\textit{last} = L'_\textit{last}$ is covered by the second case above; there cannot be valid $L$ and $L'$ such that $L_\textit{last} \leadsto L'_\textit{last}$ and $L'_\textit{last} \leadsto L_\textit{last}$ because that would induce a cycle;  $L_\textit{last} \leadsto L'_\textit{last}$ is covered by the first case above; and $L'_\textit{last} \leadsto L_\textit{last}$ is covered by the second case above. All cases are therefore covered.
				\end{proof}
			\end{proof}
			
			\step{}{\case{$L_\textit{last} \not \leq_\textit{log} L'_\textit{last} \wedge L'_\textit{last} \not \leq_\textit{log} L_\textit{last} $ (equivalent to $L_\textit{last} \parallel_\textit{log} L'_\textit{last}$)}}
			\begin{proof}
				By definition both $L$ and $L'$ are valid and non-forked, therefore the first case of $\sqcup_\mathds{L}$ applies and \texttt{Append} is called with a valid $M=L'_\textit{last}$ and returns $L''$. However, because neither $L_\textit{last} \leq_\textit{log} L'_\textit{last}$ nor $L'_\textit{last} \leq_\textit{log} L_\textit{last}$, $L_\textit{last}$ and $L'_\textit{last}$ represent different branches of a forked log. Therefore, $L''_\textit{last} = \texttt{LogPrefix}(L_\textit{last}, L'_\textit{last})$ and $L''_\textit{forks} =  \texttt{ForkProof}(L_\textit{last}, L'_\textit{last}) \neq \emptyset$. 
				
				Because $L''_\textit{forks} \neq \emptyset$, $L \leq_\mathds{L} L''$ and $L' \leq_\mathds{L} L''$. Moreover, $L''$ is now in the second phase in which a smaller $L''_\textit{last}$ (including the corresponding fork proof) results in a larger $L''$. Since \texttt{LogPrefix} is the greatest lower bound on shrinking logs (which is the reverse of computing the lower upper bound on growing logs), $L''$ is therefore the least upper bound for both $L$ and $L'$ in $(\mathds{L}(\textit{author}), \leq_\mathds{L})$. 
			\end{proof}
		
		\end{proof}

		\step{}{\case{$L_\textit{forks} \neq \emptyset \wedge L'_\textit{forks} = \emptyset$}}
		\begin{proof}
			\step{}{\case{$L_\textit{last} \leq_\textit{log} L'_\textit{last} \vee L'_\textit{last} \leq_\textit{log} L_\textit{last}$ (equivalent to $L_\textit{last} \nparallel_\textit{log}~ L'_\textit{last}$)}}
			\begin{proof}
				 $L'$ is in an earlier first phase while $L'$ is in the second phase. Since there cannot be a smaller fork than $L'$'s fork, $L'' = L \geq_\mathds{L} L'$ (greater or equal to both). There cannot be a $L'''$ smaller than $L''$ and $L$ that would also be equal to $L$. Therefore $L''$ is a least upper bound.
			\end{proof}
			
			\step{}{\case{$L_\textit{last} \parallel_\textit{log} L'_\textit{last}$}}
			\begin{proof}
				\begin{pfenum}
					\item $L''_\textit{last} = \texttt{LogPrefix}(L_\textit{last}, L'_\textit{last}) \logleadsto L_\textit{last} $ 
					\item $L''_\textit{forks} \neq \emptyset$
				\end{pfenum}
				Therefore $L'' >_\mathds{L} L$ and $L'' >_\mathds{L} L'$ (greater than both). $L''_\textit{last}$ is the greatest lower bound of $L_\textit{last}$ and $L'_\textit{last}$ because it is the result of \texttt{LogPrefix}. Therefore $L''$ is a least upper bound.
			\end{proof}
		\end{proof}
		
		\step{}{\case{$L_\textit{forks} = \emptyset \wedge L'_\textit{forks} \neq \emptyset$}}
		\begin{proof}
			Similar to the previous case, with $L$ and $L'$ exchanged.
		\end{proof}
		
		\step{}{\case{$L_\textit{forks} \neq \emptyset \wedge L'_\textit{forks} \neq \emptyset$}}
		\begin{proof}
			\step{}{\case{$L_\textit{last} \leq_\textit{log} L'_\textit{last}$}}
			\begin{proof}
				\begin{pfenum}
					\item $L''_\textit{last} = \texttt{LogPrefix}(L_\textit{last}, L'_\textit{last}) = L_\textit{last} \leq_\textit{log} L'_\textit{last}$ 
					\item $L''_\textit{forks} = L_\textit{forks}$
				\end{pfenum}
				Therefore $L'' = L$ and $L'' \geq_\mathds{L} L'$ (greater or equal to both). There cannot be a $L'''$ smaller than $L''$ and $L$ that would also be equal to $L$. Therefore $L''$ is a least upper bound.
			\end{proof}
			
			\step{}{\case{$L'_\textit{last} \leq_\textit{log} L_\textit{last}$}}
			\begin{proof}
				Similar to the previous case with $L$ and $L'$ exchanged.
			\end{proof}
			
			\step{}{\case{$L_\textit{last} \parallel_\textit{log} L'_\textit{last}$}}
			\begin{proof}
				\begin{pfenum}
					\item $L''_\textit{last} = \texttt{LogPrefix}(L_\textit{last}, L'_\textit{last}) \Rightarrow  L''_\textit{last} \logleadsto L_\textit{last} \wedge L''_\textit{last} \logleadsto L'_\textit{last} $ 
					\item $L''_\textit{forks} \neq \emptyset$
					\item $L''$ is valid (Section~\ref{sec:proofs:valid-log})
				\end{pfenum}
				Therefore $L'' >_\mathds{L} L$ and $L'' >_\mathds{L} L'$ (greater than both). $L''_\textit{last}$ is the greatest lower bound of $L_\textit{last}$ and $L'_\textit{last}$ because it is the result of \texttt{LogPrefix}. Therefore $L''$ is a least upper bound.
			\end{proof}
		\end{proof}
		\qedstep{}
		\begin{proof}
			All cases are covered.
		\end{proof}
	\end{proof}
	
	\step{}{\textbf{Monotonicity}:
	All operations that may generate a new state, when applied on log state $L$ and any possible arguments, result in a new log state either equal or larger than $L$ in $\mathds{L}(\textit{author})$ partially ordered by $\leq_\mathds{L}$.}
	\begin{proof}
	
		\step{}{\case{$L' = \texttt{Append}(L, M) \geq_\mathds{L} L$}}
		\begin{proof}

				\step{}{\case{$L_\textit{forks} = \emptyset$}}
				\begin{proof}
					\step{}{\case{$L_\textit{last} \logleadsto M$}}
					\begin{proof}
						\begin{pfenum}
							\item $L'_\textit{author} = L_\textit{author}$
							\item $L'_\textit{forks} = \emptyset$
							\item $L'_\textit{last} = M \geq_\textit{log} L_\textit{last}$
						\end{pfenum}
						Therefore, $L' \geq_\mathds{L} L$.
					\end{proof}
					
					\step{}{\case{$L_\textit{last} \geq_\textit{log} M$}}
					\begin{proof}
						$L' = L \geq_\mathds{L} L$
					\end{proof}
					
					\step{}{\case{$L_\textit{last} \parallel_\textit{log} M$}}
					\begin{proof}
						Fall through to the fork handling (Alg.~\ref{alg:log}, l.\ref{alg:log:fork-handling}).
						\begin{pfenum}
							\item $L'_\textit{author} = L_\textit{author}$
							\item $L'_\textit{forks} \neq \emptyset$
							\item $L'_\textit{last} = \texttt{LogPrefix}(L_\textit{last}, M) \logleadsto L_\textit{last} \Rightarrow L'_\textit{last} \leq_\textit{log} L_\textit{last}$
						\end{pfenum}
						Therefore, $L' \geq_\mathds{L} L$.
					\end{proof}

				\end{proof}
		
				\step{}{\case{$L_\textit{forks} \neq \emptyset$}}
				\begin{proof}
					\step{}{\case{$L_\textit{last} \nparallel_\textit{log} M$}}
					\begin{proof}
						$M$ does not provide a fork earlier on the log. We keep the previous state, therefore $L' = L \geq_\mathds{L} L$.
					\end{proof}
					
					\step{}{\case{$L_\textit{last} \parallel_\textit{log} M$}}
					\begin{proof}
						Fall through to the fork handling (Alg.~\ref{alg:log}, l.\ref{alg:log:fork-handling}).
						\begin{pfenum}
							\item $L'_\textit{author} = L_\textit{author}$
							\item $L'_\textit{forks} = L_\textit{forks} \neq \emptyset$
							\item $L'_\textit{last} = \texttt{LogPrefix}(L_\textit{last}, M) \logleadsto L_\textit{last} \Rightarrow  L'_\textit{last} \leq_\textit{log} L_\textit{last}$
						\end{pfenum}
						Therefore, $L' \geq_\mathds{L} L$.
					\end{proof}
				\end{proof}

			\qedstep{}
			\begin{proof}
				Since all cases lead to $L' = \texttt{Append}(L, M) \geq_\mathds{L} L$, \texttt{Append} is monotonic.
			\end{proof}
		\end{proof}
		
		\step{}{\case{$L'' = L \sqcup_\mathds{L} L' \Rightarrow L'' \geq_\mathds{L} L \wedge L'' \geq_\mathds{L} L'$}}
		\begin{proof}
			Since $L'' = L \sqcup_\mathds{L} L'$ computes the least upper bound of $L$ and $L'$ in $\mathds{L}(\textit{author}$ partially ordered by $\leq_\mathds{L})$, it is also monotonic.
		\end{proof}
	\end{proof}
	
	\qedstep{}
	\begin{proof}
		Because the Log satisfies the three propositions, it is a state-based CRDT.
	\end{proof}
\end{proof}

\subsubsection{Frontier}
\label{sec:proofs:convergence:frontier}

\prove{The Frontier design listed in Alg.~\ref{alg:frontier} is a state-based (convergent) CRDT.}
\begin{proof}
	\define{\begin{pfenum}
		\item $F,F' \in \mathds{F}$
		\item $L \in \mathds{L}$
	\end{pfenum}}
		
	\step{}{\textbf{Ordering}:  $\leq_\mathds{F}$ (Alg.~\ref{alg:frontier}) partially orders $\mathds{F}$.}
	\begin{proof}
		The definition of $\leq_\mathds{F}$ is analogous to that for grow-only dictionaries of counters~\cite{lavoie2023inftypset} and ledgers as grow-only dictionary of accounts~\cite{lavoie2023gocledger}: a frontier $F'$ is equal or larger than a frontier $F$ if and only if it has logs for a superset of authors, which is a partial order on the sets of authors, and each log for authors present in both $F$ and $F'$ is larger or equal in $F'$ compared to $F$, which is also a partial order, as proven in Section~\ref{sec:proofs:convergence:log}. Because the conjunction of partial orders is also a partial order~\cite{lavoie2023inftypset}, $\leq_\mathds{F}$ is a partial order over $\mathds{F}$.
	\end{proof}

	\step{convergence:lub}{\textbf{Least-Upper Bound}:	
		\prove{$F'' = F \sqcup_\mathds{F} F'$ is the LUB of $F$ and $F'$ in $\mathds{F}$ partially ordered by $\leq_\mathds{F}$.}
	}
	\begin{proof}
		The definition of $\sqcup_\mathds{F}$ is the composition of $\cup$ of grow-only sets of logs, which computes the least-upper bound on authors present in both $F$ and $F'$, and $\sqcup_\mathds{L}$, which computes the least-upper bound of two log states for the same author when present in both $F$ and $F'$. Since the composition of least upper bounds is also a least upper bound~\cite{lavoie2023gocledger}, $\sqcup_\mathds{F}$ computes the least upper bound of $F$ and $F'$.
	\end{proof}

	\step{}{\textbf{Monotonicity}:
	All operations that may generate a new state, when applied on frontier state $F$ and any possible arguments, result in a new frontier state either equal or larger than $F$ in $\mathds{F}$ partially ordered by $\leq_\mathds{F}$.}
	\begin{proof}
		\step{}{\case{$F' = \texttt{Update}(F, L) \geq_\mathds{F} F$}}
		\begin{proof}
			\step{}{\case{$\exists L' \in F : L'_\textit{author} = L_\textit{author}$}}
			\begin{proof}
				$\exists L'' \in F : L'' = L \sqcup_\mathds{L} L'$ and therefore $L'' \geq_\mathds{L} L'$. Every other logs in $F$ and $F'$ are equal, therefore $F'$ has a superset of the authors of $F$ and every log is equal or greater. Therefore $F' \geq_\mathds{F} F$.
			\end{proof}
			
			\step{}{\case{otherwise}}
			\begin{proof}
				$F'$ has one more log than $F$: it has a strict superset of authors, and every log in $F$ is equal to the log with the same author in $F'$ therefore $F' >_\mathds{F} F$.
			\end{proof}		
		\end{proof}
		
		\step{}{\case{$F'' = F \sqcup_\mathds{F} F' \Rightarrow F'' \geq_\mathds{F} \wedge~ F'' \geq_\mathds{F} F'$}}
		\begin{proof}
			This is a subset of the properties of an operation that computes a least upper bound, which was proven in step \stepref{convergence:lub}.
		\end{proof}
		
		\qedstep{}
		\begin{proof}
			All operations that modify the state of $F$ are monotonic and other operations on $F$ do not modify the state.
		\end{proof}
	\end{proof}
	
	\qedstep{}
	\begin{proof}
		The \textit{partial ordering}, the \textit{least upper bound} and the \textit{monotonicity} properties are all satisfied, therefore a frontier is a state-based CRDT.
	\end{proof}

\end{proof}

\subsection{Safety}

\subsubsection{Every log operation on a valid log results in a valid log}
\label{sec:proofs:valid-log}

\assume{\begin{pfenum}
	\item $L, L' \in \mathds{L}$ (therefore $L$ and $L'$ are valid)
	\item $M \in \mathds{M}$ (therefore $M$ is valid)
	\item $\textit{author} \in \mathds{A}$ 
\end{pfenum}}
\prove{All log state-changing operations involving  possibly $L,L'$ and $M$ as arguments result in a new log state $L''$ that is valid.}
\begin{proof}
	\step{}{\case{$L'' = \texttt{Initialize}_\mathds{L}(\textit{author})$}}
	\begin{proof}
			Log is in growing phase, meets the \textit{no forks}, \textit{consistent author}, and \textit{valid messages} properties.
		\step{}{\textit{no forks}: $L_\textit{forks} = \emptyset$}
		\begin{proof}
			By definition.
		\end{proof}
		
		\step{}{\textit{consistent author}: $L_\textit{last} = \bot$}
		\step{}{\textit{valid messages}: $L_\textit{last} = \bot$}
		\begin{proof}
		 	First message not yet set, consistent and valid according to definition.
		\end{proof}

		\qedstep{}
		\begin{proof}
			All three properties of growing logs are met, $L''$ is therefore valid.
		\end{proof}
	\end{proof}
	
	\step{}{\case{$L'' = \texttt{Append}(L, M)$}}
	\begin{proof}
		\step{}{\case{$L_\textit{forks} = \emptyset$ and $M \nparallel_\textit{log} L_\textit{last}$}}
		\begin{proof}
			$L$ is in growing phase and stays in growing phase, $L''$ must meet the \textit{no forks}, \textit{consistent author}, and \textit{valid messages} properties.		
			\step{}{\textit{consistent author}}
			\begin{proof}

					\step{}{\case{$L_\textit{last} \logleadsto M \Rightarrow L''_\textit{last} = M$}}
					\begin{proof}
						By definition of $\logleadsto$ and the \textit{single writer} property since $M$ is valid.
					\end{proof}
					
					\step{}{\case{$M \leq_\textit{log} L_\textit{last} \Rightarrow L'' = L$}}
					\begin{proof}
						Since $L$ is returned without being modified and is valid by assumption.
					\end{proof}
					
					\qedstep{}
					\begin{proof}
						This covers all possilities such that $M ~\nparallel_\textit{log}~ L_\textit{last}$.
					\end{proof}					
			\end{proof}
			
			\step{}{\textit{valid messages}}
			\begin{proof}
				Since $M ~\nparallel_\textit{log}~ L_\textit{last}$, either $L''_\textit{last} = M$ or $L''_\textit{last} = L_\textit{last}$. Since $L$ and $M$ are required to be valid in the pre-conditions of \texttt{Append} then $L''_\textit{last}$ is necessarily valid.
			\end{proof}
			
			\qedstep{}
			\begin{proof}
				\textit{no forks} is met by assumption and all cases meet the \textit{consistent author} and \textit{valid messages} properties.
			\end{proof}
		\end{proof}
		
		\step{lbl:fork-handling}{\case{$L_\textit{forks} = \emptyset$ and $M \parallel_\textit{log} L_\textit{last}$}}
		\begin{proof}
			$L$ is in growing phase but we potentially have a new proof of fork. If $M_\textit{author} \neq L_\textit{author}$, then $M$ is not a proof of fork and $L'' = L$, which is still in the growing phase and valid so $L''$ is also valid. Otherwise,  $L''$ is in shrinking phase and is the result of the fork handling (Alg.~\ref{alg:log}, l.~\ref{alg:log:fork-handling}): it must meet the FL1-FL7 properties:
				\step{}{FL1: \textit{non-empty forks}}
				\begin{proof}
					$L_\textit{forks}$ is not empty because $M \parallel_\textit{log} L_\textit{last}$ implies that there exists two $M',M''$ such that $M' \leq_\textit{log} M$ and $M'' \leq_\textit{log} L_\textit{last}$ and $M'_\textit{prev} = M''_\textit{prev}$.
				\end{proof}
				
				\step{}{FL2: \textit{valid last message}}	
				\begin{proof}
					$L'_\textit{last} = \texttt{LogPrefix}(L_\textit{last},M)$ is a prefix and any message in that prefix is valid by definition because $L$ and $M$ are valid, which implies their predecessors are valid as well.
				\end{proof}
				
				\step{}{FL3: \textit{consistent previous author}}	
				\begin{proof}
					$L'_\textit{last} = \texttt{LogPrefix}(L_\textit{last},M)$ is a prefix and any message in that prefix has a consistent author by definition because $L$ and $M$ are valid, which implies their predecessors have consistent authors.
				\end{proof} 	

				\step{}{FL4: \textit{valid forks}}
				\begin{proof}
					\texttt{ForkProof} only returns messages that are either $L_\textit{last}$, $M$, or predecessors of both. Since $L$ and $M$ are valid as required in the pre-condition of \texttt{Append}, $L_\textit{last}$, $M$ and any of their predecessors are valid. Therefore, messages in $L''_\textit{forks} = \texttt{ForkProof}(L_\textit{last}, M)$ are necessarily valid.
				\end{proof}
			
				\step{}{FL5: \textit{consistent author}}
				\begin{proof}
					$M$ is a proof of fork, then $L''_\textit{last} = \texttt{LogPrefix}(L_\textit{last}, M)$ and $L''_\textit{forks} = \texttt{ForkProof}(L_\textit{last}, M)$ and messages in both must all have consistent authors. $\texttt{LogPrefix}(L_\textit{last}, M)$ either returns $\bot$ in which case the author does not matter, or $M'$ which is either equal to $M$ or a valid predecessor ($M' \logleadsto M$) because $M$ is valid and has consistent authors, therefore $(L''_\textit{last})_\textit{author} = L_\textit{author}$. $L''_\textit{forks} = \texttt{ForkProof}(L_\textit{last}, M)$ has consistent author because all fork proofs are selected only among predecessors of $L_\textit{last}$ and $M$, which have consistent author because $L_\textit{last}$ and $M$ are valid.
				\end{proof}

				\step{}{FL6: \textit{valid proof}}
				\begin{proof}
					Because $M \parallel_\textit{log} L_\textit{last}$, then either $M_\textit{prev} = (L_\textit{last})_\textit{prev}$ or there exist some predecessor(s) of both messages for which it is true. Therefore \texttt{ForkProof} will return at least two different messages with the same predecessor.
				\end{proof}
			
				\step{}{FL7: \textit{consistent proof}}
				\begin{proof}
					By definitions of $\texttt{ForkProof}(M,M')$.
				\end{proof}

		\end{proof}
		
		\step{}{\case{$L_\textit{forks} \neq \emptyset$}}
		\begin{proof}
			$L$ is in shrinking phase, any later updates resulting in $L''$ must stay there and meet the \textit{valid forks}, \textit{consistent author}, \textit{valid proof}, and \textit{consistent proof} properties to stay valid.
			\step{}{\case{$M \nparallel_\textit{log} L_\textit{last}$}}
			\begin{proof}
				Regardless of whether $M \geq_\textit{log}~ L_\textit{last}$ or $M \logleadsto~ L_\textit{last}$, $M$ does not provide a new fork proof, therefore $L''=L$ and is valid because $L$ is valid.
			\end{proof}
			
			\step{}{\case{$M \parallel_\textit{log} L_\textit{last} \Rightarrow \texttt{LogPrefix}(M, L_\textit{last}) \logleadsto L_\textit{last}$}}
			\begin{proof}
				Because $M$ and $L_\textit{last}$ are concurrent, valid, and have consistent authors, there must exist a smaller log prefix than $L_\textit{last}$ with a fork proof. The proofs of validity properties are the same as for \stepref{lbl:fork-handling}.
			\end{proof}
		\end{proof}
	\end{proof}
	
	\step{}{\case{$L'' = L \sqcup_\mathds{L} L'$}}
	\begin{proof}
		\step{step:not-concurrent}{\case{$L'_\textit{last} \nparallel_\textit{log} L_\textit{last}$}}
		\begin{proof}
			\step{}{\case{$L_\textit{forks} = \emptyset \wedge L'_\textit{forks} = \emptyset$}}
			\begin{proof}
				\step{}{\case{$L_\textit{last} \leq_\textit{log} L'_\textit{last}$}}
				\begin{proof}
					$L'' = L'$ and $L'$ is valid, therefore $L''$ is also valid.
				\end{proof}
			
				\step{}{\case{$L'_\textit{last} \logleadsto L_\textit{last}$}}
				\begin{proof}
					$L'' = L$ and $L$ is valid, therefore $L''$ is also valid.
				\end{proof}
			\end{proof}
			
			\step{}{\case{$L_\textit{forks} \neq \emptyset \wedge L'_\textit{forks} = \emptyset$}}
			\begin{proof}
				$L'' = L$ and $L$ is valid, therefore $L''$ is also valid.
			\end{proof}
		
			\step{}{\case{$L_\textit{forks} = \emptyset \wedge L'_\textit{forks} \neq \emptyset$}}
			\begin{proof}
				$L'' = L'$ and $L'$ is valid, therefore $L''$ is also valid.
			\end{proof}
		
			\step{}{\case{$L_\textit{forks} \neq \emptyset \wedge L'_\textit{forks} \neq \emptyset$}}
			\begin{proof}
				Either $L_\textit{last} \leq_\textit{log} L'_\textit{last}$ or $L'_\textit{last} \logleadsto L_\textit{last}$, since $L_\textit{last}$ and $L'_\textit{last}$ are not concurrent (because \stepref{step:not-concurrent}). Then $L''_\textit{last}$ will be the smaller of the two and $L''_\textit{forks}$ includes the corresponding messages from either $L_\textit{forks}$ and/or $L'_\textit{forks}$. There are cases in which $L''_\textit{forks}$ may end up with more messages than $L_\textit{forks}$ or $L'_\textit{forks}$ but this does not influence the validity.
			\end{proof}
		\end{proof}

		\step{}{\case{$L'_\textit{last} \parallel_\textit{log} L_\textit{last}$}}
		\begin{proof}
			This implies that:
			\begin{pfenum}
				\item $L''_\textit{last} = \texttt{LogPrefix}(L_\textit{last},L'_\textit{last}) \wedge L''_\textit{last} \logleadsto L_\textit{last} \wedge  L''_\textit{last} \logleadsto L'_\textit{last}$
				\item $L''_\textit{forks} = \texttt{ForkProof}(L_\textit{last}, L'_\textit{last})$
			\end{pfenum}
			~\newline
			
			\step{}{FL1: \textit{non-empty forks}}
			\begin{proof}
				$L''_\textit{forks}$ is not empty because $L_\textit{last} \parallel_\textit{log} L'_\textit{last}$ implies that there exists two $M,M'$ such that $M \leq_\textit{log} L_\textit{last}$ and $M' \leq_\textit{log} L'_\textit{last}$ and $M_\textit{prev} = M'_\textit{prev}$.
			\end{proof}
				
			\step{}{FL2: \textit{valid last message}}	
			\begin{proof}
					$L''_\textit{last} = \texttt{LogPrefix}(L_\textit{last},L'_\textit{last})$ is a prefix and any message in that prefix is valid by definition because $L$ and $L'$ are valid, which implies their predecessors are valid as well.
			\end{proof}
				
			\step{}{FL3: \textit{consistent previous author}}	
			\begin{proof}
					$L''_\textit{last} = \texttt{LogPrefix}(L_\textit{last},L'_\textit{last})$ is a prefix and any message in that prefix is valid by definition because $L$ and $L'$ are valid, which implies their predecessors have consistent authors as well.
			\end{proof} 
			
			\step{}{FL4: \textit{valid forks}}
			\begin{proof}
				\texttt{ForkProof} only returns messages that are either $L_\textit{last}$, $L'_\textit{last}$, or predecessors of both. Since $L_\textit{last}$ and $L'_\textit{last}$ are valid as required in the pre-conditions, any of their predecessors are also valid. Therefore, messages in $L''_\textit{forks} = \texttt{ForkProof}(L_\textit{last}, L'_\textit{last})$ are necessarily valid.
			\end{proof}
			
			\step{}{FL5: \textit{consistent fork author}}
			\begin{proof}
				Since $L$ and $L'$ are valid, and $L_\textit{author}=L'_\textit{author}$, therefore $L_\textit{last}$ and $L'_\textit{last}$, and their predecessors have consistent authors. Since messages in $L''_\textit{forks}$ are selected from  $L_\textit{last}$ and $L'_\textit{last}$ or their predecessors, they will therefore also have a consistent author.
			\end{proof}
			
			\step{}{FL6: \textit{valid proof}}
			\begin{proof}
			Since $L_\textit{last}$ and $L'_\textit{last}$ are concurrent and valid, there must exist two different messages, one on each branch, with a shared predecessor and if so, this will be returned by \texttt{ForkProof}. Other existing proofs in $L_\textit{forks}$ and $L'_\textit{forks}$, if any, will be ignored because they are on more recent messages.
			\end{proof}
			
			\step{}{FL7: \textit{consistent proof}}
			\begin{proof}
				By definitions of $\texttt{ForkProof}(M,M')$.
			\end{proof}
		\end{proof}

	\end{proof}
\end{proof}

\subsubsection{Every frontier operation on a valid frontier results in a valid frontier}
\label{sec:proofs:valid-frontier}

\begin{proof}
	\step{}{\case{$\texttt{Initialize}_\mathds{F}$}}
	\begin{proof}
	Trivially true, because it returns an empty set with no log in it.
	\end{proof}
	
	\step{}{\case{$F'' = \texttt{Update}(F,L)$}}
	\begin{proof}
		\step{}{F1: \textit{valid logs}}
		\begin{proof}			
			\step{}{\case{$\nexists L' \in F : L'_\textit{author} = L_\textit{author}$}}
			\begin{proof}
				Because $F$ is valid, it only has valid logs and $L$  is valid (pre-conditions on \texttt{Update}). $F'' = F \cup \{ L \}$ and therefore only contains valid logs.
			\end{proof}
			
			\step{}{\case{$\exists L' \in F : L'_\textit{author} = L_\textit{author}$}}
			\begin{proof}
				Because $F$ is valid, it only has valid logs, $L$  is valid (pre-conditions on \texttt{Update}), and $F$ contains only a single log $L'$ such that $L'_\textit{author} = L_\textit{author}$. $F'' = F \backslash \{ L' \} \cup \{ L \}$: removing $L'$ from $F$ does not affect validity, and adding a valid $L$ to $F$ results in $F''$ having only valid logs.
			\end{proof}
		\end{proof}
		
		\step{}{F2: \textit{one author per frontier}}
		\begin{proof}
		When adding $L$ to $F$, if there is already an $L'$ such that $L'_\textit{author} = L_\textit{author}$, $L'$ is first removed then $L$ is added, leaving only a single log with author $L_\textit{author}$. If there are no other $L'$ such that $L'_\textit{author} = L_\textit{author}$, then adding $L$ results in having only one log with $L_\textit{author}$. Finally there can't be more than one log with the same author within $F$ because $F$ is valid.
		\end{proof}
	\end{proof}
	
	\step{}{\case{$F'' = \sqcup_\mathds{F}(F,F')$}}
	\begin{proof}
		\step{}{F1: \textit{valid logs}}
		\begin{proof}
		Because $F$ and $F'$ are valid they contain only valid logs. $F''$ is the union of all logs with authors tha are either only in $F$ or $F'$, and the merging of logs ($\sqcup_\mathds{L}$) with author that are in both $F$ and $F'$. Since the merge results in valid logs (Section~\ref{sec:proofs:valid-log}) and all other logs are valid and unchanged, then $F''$ only contains valid logs.
		\end{proof}
		
		\step{}{F2: \textit{one author per frontier}}
		\begin{proof}
		Since $F$ and $F'$ are valid they have at most one log per author in each. When merging, any two logs respectively in $F$ and $F'$ with the same author will result in a single log in $F''$. Therefore, $F''$ has at most one log per author.
		\end{proof}
	\end{proof}
	
	\qedstep
	\begin{proof}
		Since all properties of valid frontiers are maintained for all cases of all frontier operations, then every frontier operation on a valid frontier results in a valid frontier.
	\end{proof}
\end{proof}

\subsection{Liveness}

\subsubsection{All correct replicas will eventually have a shrinking log replica for every log that presented different branches of a fork to correct replicas.}
\assume{\begin{pfenum}
	\item Every correct frontier replica is transitively connected to every other correct replica.
	\item Every correct frontier replica updates its own state by merging with the latest state of any other correct frontier replica, infinitely often but with potentially arbitrary long waiting periods between merges.
\end{pfenum}}
\define{\begin{pfenum}
	\item $n$ is the number of correct replicas. 
	\item $\mathcal{R} = \{ F_i \in \mathds{F} : i \in {[}1,n{]} \} $ is the set of the (valid) frontier states of correct replicas.
	\item At some point in time, there exists $\mathds{Y} : \emptyset \subset \mathds{Y} \subseteq \mathds{A}$, for which some logs depend on forked messages from authors in $\mathds{Y}$, \textit{i.e.}, $\exists M,M' \in \mathds{M}$ such that: 
	\begin{pfenum}
		\item \textit{(valid messages)}: $M$ and $M'$ are valid;
		\item \textit{(valid fork)}: $M \neq M' \wedge M_\textit{author} = M'_\textit{author}\wedge M_\textit{prev} = M'_\textit{prev}$;
		\item \textit{(diverging logs)}:  $\exists L \in F_i \in \mathcal{R} , L' \in F_j \in \mathcal{R} : M \leq_\textit{log} L_\textit{last} \wedge M' \leq_\textit{log} L'_\textit{last}$.
	\end{pfenum}
	
\end{pfenum}}
\prove{Eventually, for all  $F_i \in \mathcal{R}$ and for each $\textit{author} \in \mathds{Y}$, $\exists$ valid $L \in F_i : L_\textit{author} = \textit{author} \wedge L_\textit{forks} \neq \emptyset$.}
\begin{proof}
The fork represented by $M$ and $M'$ is initially not replicated on any of the correct replica: some replica replicates one branch and some other another branch. Because of the assumptions all replicas will eventually replicate both branches because every frontier replica will eventually update their log replica for the same author with both branches. And because of eventual convergence, once no earlier fork is made and replicated with correct replicas, all correct replicas will agree on the state of fork logs.
\end{proof}

\section{Related Work}
\label{sec:rel-work}

In this section, we contrast our work to others and provide additional comments that were not covered in the Background (Section~\ref{sec:background}).

\subsection{Byzantine Fault-Tolerance}

The Byzantine Generals Problem~\cite{lamport2019byzantinegenerals} was introduced by \textit{Lamport et al.} as an analogy to illustrate the problem of designing algorithms in which correct processes may agree on a value even in the presence of faulty processes that may provide inconsistent messages. The Byzantine qualifier was later adopted to describe algorithms that can tolerate arbitrary behaviour from faulty processes. As later explained by Cachin~(Chapter 3, \cite{lamport2019concurrency}), the reliable dissemination of one value, solved by the core algorithms used to solve the Byzantine Generals (formalized in ~\cite{pease1980pease}), is \textit{Byzantine broadcast} and solutions using signed messages enable two correct (transitively connected) processes to reach agreement in the presence of an arbitrarily large number of Byzantine processes, because they can disseminate messages that cannot be tampered. Kleppmann and Howard~\cite{kleppmann2020bec} provide a stronger causal reliable broadcast in the same tradition. \textit{2P-BFT-Log} totally orders all sender messages from correct processes, but since Byzantine processes may violate this ordering with forks, an application that uses our logs should be able to recover from forks.

More generally, the Byzantine Generals Problem has emphasized the importance of \textit{a priori safe} decisions, in that once an attack is engaged it is not possible to undo the operation and a bad decision may lead to catastrophic outcomes to loyal generals. The generals are therefore allowed to exchange messages until they reach agreement \textit{prior} to attacking. Many problems don't require such a strong level of safety and instead inconsistency may be repaired once malicious participants successfully carried them: \textit{e.g.}, the invalidation of overspent tokens~\cite{lavoie2023gocledger} for correct participants could be compensated by a collective insurance scheme and the malicious participant blocked. For these problems, unrepudiable \textit{detection} of incorrect behaviour followed by \textit{reparation} towards honest participants is sufficient to tolerate incorrect behaviour and can possibly be cheaper than attempting complete prevention on all operations.

\subsection{Fork Consistency}

Out of over 50 different consistency models surveyed from distributed systems, and storage systems research~\cite{viotti2016consistency-db-survey}, the closest work to ours is the \textit{fork consistency} model~\cite{mazieres2002sfs-byzantine-storage}. In this model, an adversarial server implementing a file system may present inconsistent operations to different clients but once it does, clients are partitioned into groups and member of different groups may never see other groups subsequent operations. However, this model and a later refinement~\cite{li2007beyondonethirdbft} do not implement eventual consistency between replicas because it does not specify what clients should do once forks have been observed. In contrast, our \textit{2P-BFT-Log} design provides \textit{strong eventual consistency}~\cite{Shapiro2011CRDTs} because all replicas will eventually agree on the greatest lower bound between all forks as the latest non-forked message of every log. 

Depot~\cite{mahajan2011depot} describes at a high-level a client-server Cloud Storage system that uses append-only logs with fork detection and recovery in a model they call Fork-Join-Causal-Consistency. After a fork, correct replicas will still accept updates from forked logs as long as the fork has been vouched by a correct replica. In contrast, \textit{2P-BFT-Log} is a replicated data structure that can be used in both peer-to-peer and client-server environments and we precisely described all core algorithms.  Moreover, in \textit{2P-BFT-Log} forks are handled in an explicit second \textit{shrinking} phase that bounds the number of possible new forks to at most the remaining number of entries in the log.

\subsection{Timeline Entanglement and Causal Histories}

 Our append-only logs augmented with dependencies are similar to secure timelines~\cite{maniatis2002secure-timeline-entanglement} but provide a complete specification of the data structure behaviour after a fork is discovered, with the novel contribution of an eventually-consistent shrinking phase. Our append-only logs can also be seen as a reification a reification of \textit{causal histories}~\cite{schwarz1994detecting} for correct authors.

\section{Conclusion}
\label{sec:conclusion}

We have presented \textit{2P-BFT-Log}, a two-phase Byzantine Fault-Tolerant single-author append-only log design as a state-based Conflict Free Replicated Datatype. The key idea and novel contribution of our design is to add a \textit{shrinking} phase after the usual growing phase of append-only log, that is triggered by the discovery of a fork, and provides eventual consistency between all correct replica on the greatest lower bound to all forks known by correct replicas.

Our design enables establishing a total order between the messages of correct authors while providing eventual detection of concurrent messages for malicious authors. This enables, in the context of an accounting application for example, to ensure correct authors maintain non-negative balances while malicious authors double-spending, which can only be done with concurrent messages, is eventually detected.

Moreover, our design provides eventual consistency on the set of messages that affect correct authors' logs, because these are explicitly listed as dependencies. Our design also limits the window of opportunity after an initial attack is carried and guarantees that forked logs become dead, i.e. they cannot be extended with any more messages through correct replicas, once the fork proofs have been replicated by all correct replicas.

However, our design does not solve the issue of how correct authors may repair the damage done by malicious authors with concurrent messages. This will be the focus of future work, including but not limited to, how correct authors may recover from double-spent tokens. We believe many existing distributed algorithms that were originally designed to prevent bad outcomes, might be adapted instead to recover from malicious behaviour, which might well be significantly cheaper.

\section{Acknowledgements}
\label{sec:acknowledgements}

We thank Prof. Christian F. Tschudin for fostering a research environment allowing detours and playfulness in the process, as well as providing financial support for this work and feedback on earlier versions of this work. 

We would also like to thank the Secure-Scuttlebutt community for its enthusiasm in general, being a great springboard for discussions that help identify important and practically relevant underlying technical problems, and its relevant and timely discussions on our papers. In particular, we would like to thank Aljoscha Meyer for feedback on a previous version of this paper.

We also thank the Swiss tax payers for contributing their hard-earned funds to make a Swiss academia possible in general, and this paper in particular. We hope our contribution to knowledge will provide general value in different forms many times larger than what it cost you.

We would also finally like to thank you, the reader, to have made it to the end of this paper. If the ideas in this paper have been of any use, we would like to hear from you. Academia is mostly geared to track citations by other papers and the prestige of conference and journals in which papers were published, so it can be hard to assess impact beyond these two metrics. If you take a few minutes to send us an email, we will have a better idea of how useful these ideas have been outside of academia.

\newpage

\bibliographystyle{plainurl}
\bibliography{main}

\appendix

\newpage
\section{Notation and Conventions}
\label{apdx:notation}

We use notations and conventions that are good graphic mnemonics for the concepts and make the algorithms easier to reason about in the proofs. The semantics are:

\begin{itemize}
	\item \textbf{Variables} are written in \textit{italic}:
		\begin{itemize}
			\item  lower case when containing literal values, ex: $id$;
			\item upper case when containing an object with multiple fields, a dictionary with multiple key-value pairs, or a set with multiple elements. For example, $A$ for account object, $L$ for a ledger dictionary, and $S$ for a set;
		\end{itemize}
	\item An \textbf{object's field} is accessed using a subscript, ex: $A$'s identifier stored in field $\textbf{id}$ is accessed $A_{\scriptsize \textbf{id}}$;
	\item An empty \textbf{dictionary} is written $\{\}$, accessing a dictionary $L$'s value stored under key $id$ is written $L[id]$, accessing all the keys of $L$ is written $L_*$;
	\item \textbf{Assigning} a new value to a variable, a field, or a dictionary entry uses $\leftarrow$, ex: $id \leftarrow id'$, $A_{\scriptsize\textbf{id}} \leftarrow id$, $L[id] \leftarrow A$. Variables, fields, and dictionaries are mutable and can be modified in place;
	 \item A \textbf{key-value} pair for dictionaries is written $\textit{key} \rightarrow \textit{value}$;
	 \item We use "dictionary-comprehension", similar to Python, for inline initialization fo dictionares, ex: ${ \textit{key} \rightarrow \textit{value} ~\textbf{for}~ \textit{key} ~\textbf{in}~ K }$;
	\item \textbf{Different states} for replicas of objects or dictionaries are written with $'$ and $''$ using the same variable name, ex: $A, A', A''$. We represent output values of functions using variable names with $'$ or $''$ to show they are later states of the same replica;
\end{itemize}

Apart from these, we use common mathematical and pseudo-code conventions: 
\begin{itemize}
	\item $x \oplus y$ is the concatenation of the string representation of $x$ and $y$
	\item $x \in X$ is an element $x$ in a set $X$ and $x \notin X$ means $x$ is not in a set $X$;
	\item $X \subseteq Y$ means $X$ is a subset of $Y$ which may include up to all elements of $Y$;
	 \item $\sum$ is a summation;
	 \item $\sum\limits_{x \in X} x$ is the sum of all elements in $X$;
	 \item $\leq$ is smaller or equal;
	 \item $\bigwedge$ and $\textbf{and}$ both represent a logical \textit{and};
	 \item $\bigwedge\limits_{x \in X} x$ is the logical and between all elements in $X$;
	  \item $\bigcup\limits_{x \in X} \{ x \}$ is the union of all elements in $X$;
	 \item $\textbf{for}~x ~\textbf{in}~ X~\textbf{do}$ iterates over all values in $X$ sequentially assigning them to $x$.
\end{itemize}

\end{document}